\documentclass[%
 reprint,
superscriptaddress,
 amsmath,amssymb,
 aps,
pra,
]{revtex4-2}

\usepackage{graphicx}
\usepackage{dcolumn}
\usepackage{bm}
\usepackage[unicode]{hyperref}
\newcommand{\beq}{\begin{equation}}
\newcommand{\eeq}{\end{equation}}

\newcommand{\vect}[1]{\mathbf{#1}}
\newcommand{\Eq}[1]{Eq.~\eqref{#1}}

\DeclareMathOperator{\im}{Im}

\newcommand{\I}{\mathrm{i}}

\newcommand{\D}{\mathrm{d}}
\newcommand{\nn}{\nonumber}
\begin{document}

\preprint{APS/123-QED}

\title{Scattering Hypervolume of Spin-Polarized Fermions}

\author{Zipeng Wang}
\author{Shina Tan}%
 \email{shinatan@pku.edu.cn}
\affiliation{%
International Center for Quantum Materials, Peking University, Beijing 100871, China 
}%


\date{\today}

\begin{abstract}
We analyze the collision of three identical spin-polarized fermions at zero collision energy, assuming arbitrary finite-range potentials,
and define the corresponding three-body scattering hypervolume $D_F$.
The scattering hypervolume $D$ was first defined for identical bosons in 2008 by one of us. It is the three-body analog of the two-body scattering length. We solve the three-body Schr\"{o}dinger equation asymptotically when the three fermions are far apart or one pair and the third
fermion are far apart, deriving two asymptotic expansions of the wave function.
Unlike the case of bosons for which $D$ has the dimension of length to the fourth power,
here the $D_F$ we define has the dimension of length to the eighth power.
We then analyze the interaction energy of three such fermions with momenta $\hbar\vect{k}_1$, $\hbar\vect{k}_2$ and $\hbar\vect{k}_3$ in a large periodic cubic box. The energy shift due to $D_F$ is proportional to $D_F/\Omega^2$, where $\Omega$ is the volume of the box. We also calculate the shifts of energy and pressure of spin-polarized Fermi gases due to a nonzero $D_F$ and the three-body recombination rate of spin-polarized ultracold atomic Fermi gases at finite temperatures.

\end{abstract}

\maketitle


\section{\label{sec:level1}Introduction}
When electrically neutral particles collide with small energies, such that the de Broglie wave lengths are large compared to the range of interaction, their effective interactions can be characterized by a
small number of parameters such as the two-body $s$-wave scattering length. For identical spin-polarized fermions, however, the $s$-wave collision is forbidden due to the Pauli exclusion principle, and the low-energy effective interaction is dominated by the $p$-wave scattering volume $a_p$.
All the two-body effective parameters for the interaction can be extracted from the wave functions
for the two-body collision at energies equal to or close to zero, outside of the physical range of the
interaction potential \cite{tan2008three}. The $p$-wave scattering volume $a_p$, for example, can be extracted from
the wave function of the two fermions colliding at zero incoming kinetic energy: 
\begin{equation}\label{phi2}
\phi^{(1,m)}(\vect{s})=\left(\frac{s}{3}-\frac{a_p}{s^2}\right)\sqrt{\frac{4\pi}{3}}Y_1^{m}(\hat{\vect s}),~~\text{if }s>r_e,
\end{equation}
where $\vect s$ is the spatial vector extending from one fermion to the other, $r_e$ is the range of the interaction potential,
$Y_l^m(\hat{\vect s})$ is the spherical harmonic ($m=-l,-l+1,\dots,l$ is the magnetic quantum number).

The effective three-body interaction at small collision energies can also be described by some effective parameters, such as the scattering hypervolume $D$ which was first defined for three identical bosons by one of us \cite{tan2008three}. It is the three-body analog of the two-body $s$-wave scattering length $a$, and is a fundamental parameter determining the effective strength of three-body interactions at small collision energies.
$D$ affects the energies of dilute Bose-Einstein condensates \cite{tan2008three}.
The three-body recombination \cite{moerdijk1996decay,fedichev1996three,esry1999recombination,nielsen1999low,bedaque2000three,braaten2001three,HammerRevModPhys.85.197} rate is proportional to the imaginary part of $D$ \cite{zhu2017three,braaten2006universality}.
$D$ determines the effective three-body coupling constant in the effective-field theoretical description of low energy particles \cite{tan2008three,braaten1999quantum,HammerRevModPhys.85.197}.
The value of $D$ has been numerically computed for identical bosons interacting with hard-sphere \cite{tan2008three}, Gaussian \cite{zhu2017three}, square-well \cite{mestrom2019scattering} and Lennard-Jones \cite{mestrom2020van} potentials.
Recently, the definition of scattering hypervolume is generalized to three particles with unequal masses \cite{wang2021threebody}, two identical bosons and a particle with a different mass \cite{Mestrom2021pwave}, and three identical spin-1 bosons \cite{mestrom2021threebody}. In these three-body systems, the dimensions of the corresponding scattering hypervolumes are all $[\mathrm{length}]^4$. 

Can we also define a scattering hypervolume for three identical fermions in the same spin state?
In this paper, we study the zero energy collison of three spin-polarized fermions with total orbital angular momentum $L=1$, assuming arbitrary finite-range potentials. We solve the three-body Schr\"{o}dinger equation asymptotically and get two expansions for the three-body wave function $\Psi$, one of which is applicable when all the three fermions are far away from each other and is named 111 expansion, the other of which is applicable when two fermions are held at a fixed distance and the third fermion is far away from the two and is named 21 expansion. The new scattering hypervolume $D_F$ is the dominant three-body parameter in these expansions. 
The dimension of $D_F$ is $[\mathrm{length}]^8$. The parameter $D_F$ is applicable only if the thermal de Broglie wavelengths of the fermions are much larger than $r_e$, namely if the temperature $T\ll T_e$, where
\beq
T_e\equiv\frac{\hbar^2}{2M_Fr_e^2k_B}.
\eeq
Here $\hbar$ is Planck's constant over $2\pi$, $M_F$ is the mass of each fermion, and $k_B$ is the Boltzmann constant.

In Section~\ref{sec:asymp} of this paper, we derive the 111 expansion and the 21 expansion for the collision of three identical spin-polarized fermions at zero energy. We assume that the fermions are electrically neutral, and the interaction potentials are finite-ranged, vanishing beyond a certain range $r_e$.
In this paper we will expand the three-body wave function $\Psi$ to the order $B^{-6}$ in the 111 expansion and to the order $R^{-7}$ in the 21 expansion, which are the orders at which the three-body scattering hypervolume $D_F$ first appears.
Here $B$ is the hyperradius, defined as the square root of a half of the sum of the squares of the three interfermionic distances;
see \Eq{hyperradius} below. $R$ is the distance between the center of mass of the two fermions held at a fixed distance and
the third fermion which is far away from the two; see \Eq{Ri} below.

In Section~\ref{sec:energy} we calculate the shift of the energy of three fermions in a large periodic cubic box due to a nonzero $D_F$. 
We then further calculate the shifts of energy and pressure caused by $D_F$ for a homogeneous spin-polarized Fermi gas at finite temperatures.

In Section~\ref{sec:recombination} we derive the formula for the three-body recombination rate in spin-polarized ultracold atomic Fermi gases. We find that in an intermediate temperature regime $T_F\ll T\ll T_e$ (where $T_F$ is the Fermi temperature), $\D n/\D t$ is proportional to $n^3T^2$, in agreement with Refs.~\cite{PhysRevA.65.010705,PhysRevLett.120.133401,top2020spinpolarized}, while at low temperatures, $T\ll T_F$, $\D n/\D t$ is proportional to $n^{13/3}$.

\section{ASYMPTOTICS OF THE THREE-BODY WAVE FUNCTION\label{sec:asymp}}
We consider identical spin-polarized fermions with mass $M_F$ each.
We assume that the interactions among these fermions are finite-ranged and depend only on the interparticle distances, and thus they are invariant under translation, rotation and Galilean transformations.

If the three fermions collide with zero energy,
the three-body wave function $\Psi$ satisfies the Schr\"{o}dinger equation:
\begin{equation}\label{3body_equ}
\begin{split}
&\Big[-\frac{\hbar^2}{2M_F}(\nabla_1^2+\nabla_2^2+\nabla_3^2)+V(s_1)+V(s_2)\\
&\quad+V(s_3)+V_{123}(s_1,s_2,s_3)\Big]\Psi(\vect r_1,\vect r_2,\vect r_3)=0,
\end{split}
\end{equation}
where $\vect r_i$ is the position vector of the $i$th fermion, and
\beq
\vect s_i\equiv\vect r_j-\vect r_k.\label{s_i}
\eeq
In the above equation and in the following, $(i,j,k)=(1,2,3)$, $(2,3,1)$, or $(3,1,2)$.
$V(s_i)$ is the interaction potential between the $j$th fermion and the $k$th fermion, and $V_{123}(s_1,s_2,s_3)$ is the three-body potential.
We assume the total momentum of three fermions is zero (which means we study the problem in the center-of-mass frame), and thus $\Psi$ is translationally invariant:
\beq\label{3body_translational}
\Psi(\vect r_1+\delta\vect r,\vect r_2+\delta\vect r,\vect r_3+\delta\vect r)=
\Psi(\vect r_1,\vect r_2,\vect r_3)
\eeq
for any $\delta\vect r$. 

Equations~\eqref{3body_equ} and \eqref{3body_translational} do not uniquely determine the wave function for the zero energy collision.
We need to also specify the asymptotic behaviour of $\Psi$ when the three fermions are far apart. The leading-order term in $\Psi$ when
$s_1,s_2,s_3$ go to infinity simultaneously, $\Psi_0$, should satisfying the Laplace equation $(\nabla_1^2+\nabla_2^2+\nabla_3^2)\Psi_0=0$,
and scale like $B^p$ at large $B$, where
\begin{equation}\label{hyperradius}
B\equiv\sqrt{\left(s_1^2+s_2^2+s_3^2\right)/2}
\end{equation}
is the hyperradius. The most important three-body wave function for zero-energy collisions, for purposes of understanding ultracold collsions, should be the one with the minimum value of $p$. The larger the value of $p$, the less likely it is for the three particles to come to the range of interaction
within which they can interact. (For this same reason, in the study of two-body ultracold collisions of identical fermions, the $p$-wave collision is usually the most important one.) One can easily show that the minimum value of $p$ for three identical spin-polarized fermions in three spatial dimensions is $2$.
There are only three linearly independent three-body wave functions for the zero-energy collision with $p=2$, and they all have total orbital angular momentum quantum number $L=1$, and they form an irreducible representation of the rotational group, and can be distinguished using the magnetic quantum number $M=-1,0,1$. These three-body wave functions are denoted as $\Psi_1^M$.

For later use, we define the Jacobi coordinates \cite{braaten2006universality,nielsen2001three} used in this paper. 
$\vect{s}_i$ has been defined in \Eq{s_i}. We define $\vect{R}_i$ as the vector extending from the center of mass of the $j$th fermion and the $k$th fermion to the $i$th fermion:
\begin{equation}\label{Ri}
\vect{R}_i\equiv\vect{r}_i-(\vect{r}_{j}+\vect{r}_{k})/2.
\end{equation}
We also define three hyperangles:
\begin{equation}
\theta_i\equiv\arctan\frac{2R_i}{\sqrt{3} s_i}.
\end{equation}
$s_i$, $R_i$, $\theta_i$ and $B$ satisfy the following relations:
\begin{equation}
s_i=\frac{2}{\sqrt{3}} B \cos\theta_i,\quad R_i= B \sin \theta_i.
\end{equation}
The $\vect s_i$, $\vect R_i$, $B$, and $\theta_i$ defined above are the same as the corresponding variables defined for identical bosons in Ref.~\cite{tan2008three}.

\subsection{Two-body special functions}
We define the
two-body special functions $\phi^{(l,m)}(\vect{s})$, $f^{(l,m)}(\vect s)$, $g^{(l,m)}(\vect s)$, \dots, for the collision
of two particles with orbital angular momentum quantum number $l$ and  magnetic quantum number $m$ along the $z$ direction \cite{tan2008three,wang2021threebody}:
\begin{subequations}\label{Hphi,Hf,Hg}
\begin{align}
 &\widetilde{H}\phi^{(l,m)}=0,\\
 &\widetilde{H}f^{(l,m)}=\phi^{(l,m)},\\
 &\widetilde{H}g^{(l,m)}=f^{(l,m)},\\
 &\dots\nonumber
\end{align}
\end{subequations}
where $\hbar^2\widetilde{H}/M_F$ is the two-body Hamiltonian for the collision of two fermions in the center-of-mass frame, and
\begin{equation}
    \widetilde{H} \equiv -\nabla_{\vect s}^2+ \frac{M_F}{\hbar^2} V(s).
\end{equation}
For identical spin-polarized fermions, $l$ must be odd due to Pauli principle. We use symbols $p,f,h,\cdots$ to represent $l=1,3,5,\cdots$.

Given the two-body special functions $\phi^{(l,m)}$, $f^{(l,m)}$, $g^{(l,m)}$, \dots, we can express the wave function
for the collision of two particles at any small nonzero energy $E=\hbar^2 k^2/M_F$ as an infinite series in $k^2$ \cite{tan2008three,wang2021threebody}:
\begin{equation}\label{two-body-nonzero-E:expand}
\phi^{(l,m)}_k(\vect{s})=\phi ^{(l,m)}(\vect{s})+k^2 f^{(l,m)}(\vect{s})+ k^4 g^{(l,m)}(\vect{s})+\cdots.
\end{equation}

To complete the definition of $\phi^{(l,m)}$, we need to specify its overall amplitude. Since the potential $V(s)$ vanishes beyond a finite range $r_e$, $\phi^{(l,m)}$ takes a simple form at $s>r_e$:
\begin{equation}\label{two-body-phi}
\phi^{(l,m)}(\vect{s})=\left[ \frac{s^l}{(2l+1)!!}-\frac{(2l-1)!!a_l}{s^{l+1}}\right]\sqrt{\frac{4\pi}{2l+1}}Y_l^m (\hat{\vect{s}}),
\end{equation}
where $Y_l^m$ is the spherical harmonic, and $a_l$ is the two-body $l$-wave scattering volume
(with dimension $[\mathrm{length}]^{2l+1}$). We have fixed the overall amplitude of $\phi^{(l,m)}$ by specifying the coefficient of the term $\propto s^l$.

The solution to the equation $\widetilde{H}f^{(l,m)}=\phi^{(l,m)}$ is not unique, because if $f^{(l,m)}$ satisfies this equation, then $f^{(l,m)}+\text{(arbitrary coefficient)}\times\phi^{(l,m)}$ also satisfies this equation.
To complete the definition of $f^{(l,m)}$, we specify that in the expansion of $f^{(l,m)}(\vect{s})$ at $s>r_e$ we do \emph{not} have
the term $\propto s^{-l-1}$ (if such a term exists, we can add a suitable coefficient times $\phi^{(l,m)}(\vect{s})$ to $f^{(l,m)}(\vect{s})$ to cancel this term). Then at $s>r_e$ we have the following analytical formula for $f^{(l,m)}(\vect{s})$:
\begin{equation}\label{two-body-f}
\begin{split}
&f^{(l,m)}(\vect{s})=\bigg[-\frac{s^{l+2}}{2(2l+3)!!}-\frac{a_l r_l s^l}{2(2l+1)!!}\\
&\mspace{29mu}\quad\quad\quad\quad-\frac{(2l-3)!!}{2}a_{l}s^{1-l}\bigg]\sqrt{\frac{4\pi}{2l+1}}Y_l^m (\hat{\vect{s}}).
\end{split}
\end{equation}
For brevity we do not show the explicit formula for $g^{(l,m)}$ as it is not used in this paper.

The two-body special functions will appear in the 21 expansions of the three-body wave functions at zero collision energy.

If the magnetic quantum number $m=0$, these two-body functions we have defined here are the same as the special functions defined in Ref.~\cite{wang2021threebody} if one sets $\hat{\vect n}=\hat{\vect z}$ in Ref.~\cite{wang2021threebody}. 

One can show \cite{tan2008three,wang2021threebody} that $a_{l}$ which first appears in \Eq{two-body-phi} is the two-body $l$-wave scattering volume, $r_{l}$ which first appears in \Eq{two-body-f} is the two-body $l$-wave effective range, and they are related to the scattering phase shift $\delta_l(k)$ in the well-known effective range expansion \cite{hammer2010causality,hammer2009}:
\begin{equation}\label{effective-range-expansion}
k^{2l+1}\cot \delta_{l}(k)=-\frac{1}{a_{l}}+\frac{1}{2}r_{l} k^2+O(k^4).
\end{equation}

\subsection{111 expansion and 21 expansion}
As in our previous works \cite{tan2008three,wang2021threebody}, we derive two asymptotic expansions for the three-body wave function $\Psi_1^M$.
When the three fermions are all far apart from each other, such that the pairwise distances $s_1$, $s_2$, $s_3$ go to infinity simultaneously for any fixed ratio $s_1:s_2:s_3$, we expand $\Psi_1^M$ in powers of $1/B$, and this expansion is called the 111 expansion.
When one fermion (the $i$th fermion) is far away from the other two (the $j$th and the $k$th fermions), but the two fermions ($j$ and $k$) are held at a fixed distance $s_i$, we expand $\Psi_1^M$ in powers of $1/R_i$, and this is called the 21 expansion. The two expansions are
\begin{subequations}
\begin{align}
    &\Psi_1^M=\sum_{p=-2}^{\infty} \mathcal{T}^{(-p)}(\vect{r}_1,\vect{r}_2,\vect{r}_3),\label{111-form}\\
    &\Psi_1^M=\sum_{q=-1}^{\infty}\mathcal{S}^{(-q)}(\vect{R},\vect{s}),\label{21-form}
\end{align}
\end{subequations}
where $\mathcal{T}^{(-p)}$ scales as $B^{-p}$, $\mathcal{S}^{(-q)}$ scales as $R^{-q}$. Without loss of generality, here we suppose the Jacobi coordinates $\vect{s}=\vect{s}_1$, and $\vect{R}=\vect{R}_1$.

$\mathcal{T}^{(-p)}$ satisfies the free Schr\"odinger equation outside of the interaction range:
\beq\label{free Schrodinger}
-\frac{\hbar^2}{2M_F}\Big(\nabla_1^2+\nabla_2^2+\nabla_3^2\Big)\mathcal{T}^{(-p)}=0.
\eeq
If one fermion is far away from the other two, \Eq{3body_equ} becomes
\begin{equation}
\Big[-\frac{\hbar^2}{M_F}\nabla_{\vect{s}}^2+V(s)-\frac{3\hbar^2}{4M_F}\nabla_{\vect{R}}^2\Big]\Psi_1^M =0.
\end{equation}
Therefore, $\mathcal{S}^{(-q)}$ satisfies the following equations,
\begin{equation}
\begin{split}
    &\widetilde{H} \mathcal{S}^{(1)}=0,\quad\widetilde{H} \mathcal{S}^{(0)}=0,\\
    &\widetilde{H} \mathcal{S}^{(-q)}=\frac{3}{4}\nabla_{\vect{R}}^2 \mathcal{S} ^{(-q+2)}\quad (q\geq 1).
\end{split}
\end{equation}

If $s\ll R$, we can further expand $\mathcal{T}^{(-p)}$ as
\beq
\mathcal{T}^{(-p)}=\sum_{i} t^{(i,-p-i)},
\eeq
where $t^{(i,j)}$ scales like $R^i s^j$.
If $s\gg r_e$, we can further expand $\mathcal{S}^{(-q)}$ as
\beq
\mathcal{S}^{(-q)}=\sum_{j} t^{(-q,j)}.
\eeq
Because the three-body wave function $\Psi_1^M$ may be expanded as $\sum_p \mathcal{T}^{(-p)}$ at $B\to\infty$,
and may also be expanded as $\sum_q \mathcal{S}^{(-q)}$ at $R\to\infty$, the $t^{(i,j)}$ in the above two expansions
should be the same. In fact the wave function has a double expansion $\Psi_1^M=\sum_{i,j}t^{(i,j)}$ in the region $r_e\ll s\ll R$.

\begin{figure}[htb]
\includegraphics[width=0.5\textwidth,height=0.5\textwidth]{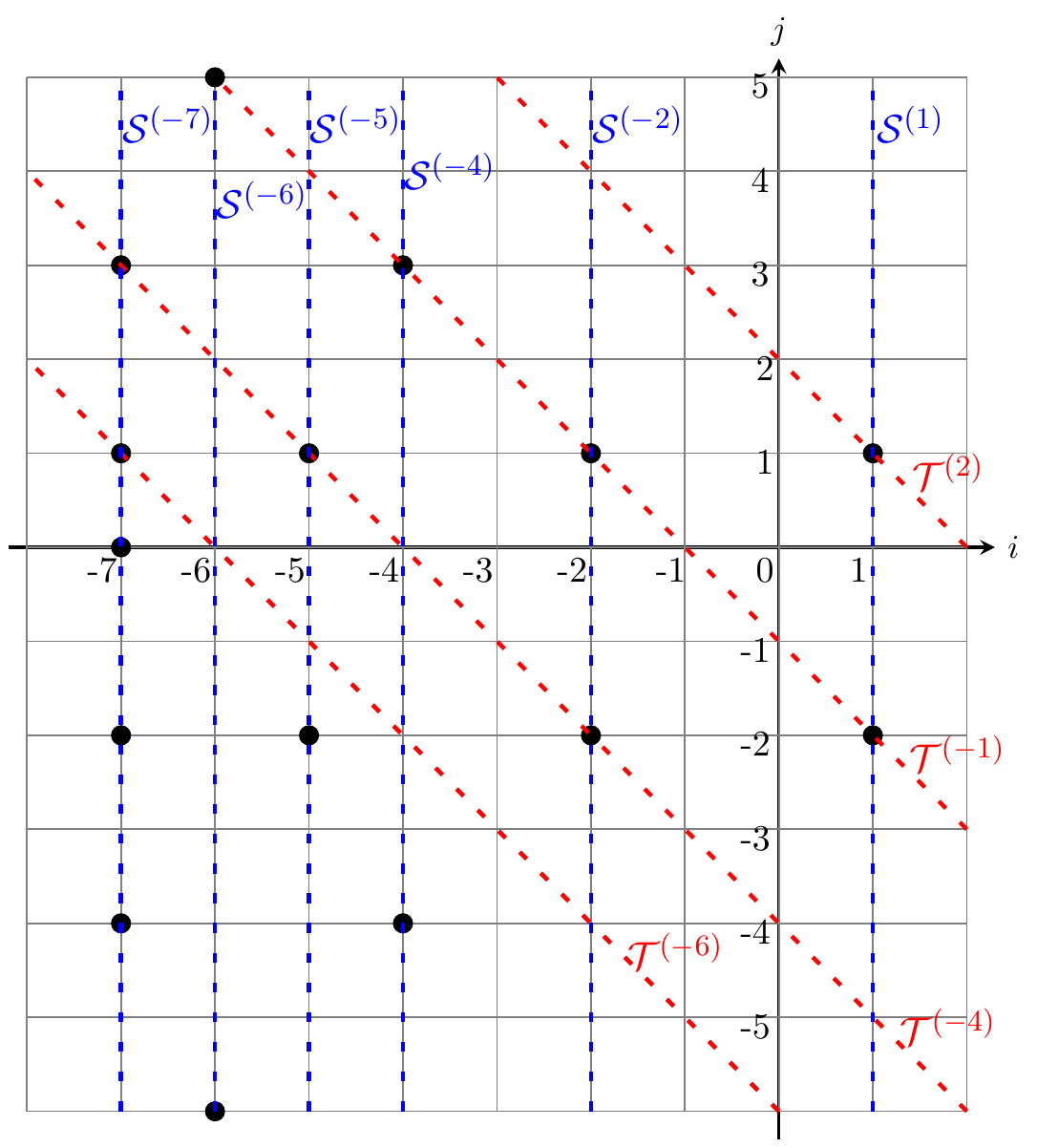}
\caption{\label{fig:expansion} Diagram of the points representing $t^{(i,j)}$ on the $(i,j)$ plane. Each point with coordinates $(i,j)$ represents $t^{(i,j)}$ which scales like $R^i s^j$. Thick dots represent those points at which $t^{(i,j)}\ne0$.
Each nonzero term $\mathcal{T}^{(-p)}$ in the 111-expansion is represented by a red dashed line satisfying the equation $i+j=-p$. For $s>r_e$, the term $\mathcal{S}^{(-q)}$ in the 21-expansion is represented by the vertical blue line satisfying the equation $i=-q$.
We have not derived the expressions for $\mathcal{T}^{(-7)}, \mathcal{T}^{(-8)}, \mathcal{T}^{(-9)}$ etc, and so the red dashed lines
corresponding to them are not shown.}
\end{figure}

We show the points at which $t^{(i,j)}\ne0$ on the $(i,j)$ plane in Fig. \ref{fig:expansion}. $\mathcal{T}^{(-p)}$ corresponds to 
the straight line with slope equal to $-1$ and intercept equal to $-p$. $\mathcal{S}^{(-q)}$ corresponds to the vertical line $i=-q$. 
Therefore, all the points $t^{(i,j)}$ satisfying $i+j=-p$ are on the line corresponding to $\mathcal{T}^{(-p)}$, and all the points $t^{(-q,j)}$ are on the line of $\mathcal{S}^{(-q)}$.

To derive the two expansions, we start from the leading-order term in the 111 expansion (which fixes the overall ampltude of $\Psi_1^M$):
\beq
\mathcal{T}^{(2)}=
Q_1^{M}(\vect{s}\times\vect{R}),~~M=-1,0,1,
\eeq
where $Q_L^M(\vect{u})$ is the harmonic polynomial,
\begin{equation}
Q_L^M(\vect{u})\equiv\sqrt{\frac{4\pi}{2L+1}}\,u^L Y_{L}^M (\hat{\vect{u}}).
\end{equation}
More explicitly, for $M=0$,
\begin{equation}
\mathcal{T}^{(2)}=A_z,
\end{equation}
and for $M=\pm1$,
\beq
\mathcal{T}^{(2)}=-\frac{M}{\sqrt2}(A_x+\I M A_y),
\eeq
where
\beq
\vect A=\vect s\times\vect R=-(\vect{r}_1\times\vect{r}_2+\vect{r}_2\times\vect{r}_3+\vect{r}_3\times\vect{r}_1)
\eeq
is a vector perpendicular to the plane of the triangle formed by the three fermions,
and $A$ is equal to twice the area of such triangle.
One can check the leading-order term $\mathcal{T}^{(2)}$ satisfies the free Schr\"odinger equation. It is also translationally invariant, and is antisymmetric under the exchange of the fermions. $\mathcal{T}^{(2)}=t^{(1,1)}$ is denoted by the point with coordinates $(1,1)$ in Fig. \ref{fig:expansion}.

We then first derive $\mathcal{S}^{(1)}$, and then derive $\mathcal{T}^{(1)}$, and then derive $\mathcal{S}^{(0)}$, and then derive $\mathcal{T}^{(0)}$,
and so on, all the way until $\mathcal{S}^{(-7)}$.
At every step, we require the 111 expansion and the 21 expansion to be consistent in the region $r_e\ll s\ll R$. See Appendices \ref{Derivation} and \ref{sec:M=pm1} for more details.

Our resultant 111 expansion is
\begin{widetext}
\begin{equation}\label{111}
\Psi_{1}^M=Q_1^{M}(\vect{s}\times\vect{R})\cdot 
\left[ 1-3 a_p\sum_{i=1}^{3} \frac{1}{s_i^3}+\frac{36a_p^2}{\pi}\sum_{i=1}^{3}\frac{\left( \theta_i-\frac{1}{4}\sin 4\theta_i\right) }{R_i^3s_i^3}-\frac{9\sqrt{3}D_F}{4\pi^3 B^8}\right] +O(B^{-7}),
\end{equation}
\end{widetext}
where $D_F$ is the three-body scattering hypervolume of identical spin-polarized fermions, and it appears at the order of $B^{-6}$. We have chosen the coefficient $-9\sqrt{3}/4\pi^3$ in front of $D_F$ to simplify the formula for the energy shift
of three identical fermions in a large periodic volume due to the three-body parameter; see Sec.~\ref{subsec:three fermions in box} for details.

Our resultant 21 expansion is
\begin{widetext}
\begin{equation}
\begin{split}
\Psi_1^M=&6\I\sqrt{\frac{2\pi}{3}} \left[ R-\frac{6a_p}{R^2}+\frac{12a_p^2}{R^5}\left( 8-\frac{9\sqrt{3}}{\pi}\right) -\frac{\xi}{R^7}\right] \sum_m C^{1,M}_{1,M-m;1,m} Y_1^{M-m}(\hat{\vect{R}})\phi^{(1,m)}(\vect{s}) \\
&+\I\left[ -\frac{630\sqrt{\pi}a_p}{R^4}+\frac{4032\sqrt{\pi}a_p^2}{R^7}\left( 16-\frac{27\sqrt{3}}{\pi}\right)\right] \sum_m C^{1,M}_{3,M-m;3,m} Y_3^{M-m}(\hat{\vect{R}})\phi^{(3,m)}(\vect{s})\\
&-\I\frac{31185\sqrt{10\pi} a_p}{4R^6}\sum_m C^{1,M}_{5,M-m;5,m} Y_5^{M-m}(\hat{\vect{R}})\phi^{(5,m)}(\vect{s})\\
&+\I\frac{324\sqrt{6\pi} a_p^2}{R^7} \left( 8-\frac{9\sqrt{3}}{\pi}\right) \sum_m C^{1,M}_{1,M-m;1,m}  Y_1^{M-m}(\hat{\vect{R}}) f^{(1,m)}(\vect{s})+O(R^{-8}),
\end{split}
\end{equation}
\end{widetext}
where
\begin{equation}
C^{J,M}_{l_1m_1;l_2m_2}=\left\langle l_1,m_1;l_2,m_2|J,M\right\rangle
\end{equation}
is the Clebsch-Gordan coefficient, and $\xi$ is a parameter related to $D_F$,
\begin{equation}
\xi=\frac{9\sqrt{3}D_F}{4\pi^3}-81a_p^3 r_p\left( 8-\frac{9\sqrt{3}}{\pi}\right) .
\end{equation}


\section{Energy shifts and thermodynamic properties}\label{sec:energy}
In this section, we study the energy shifts of $N$ identical spin-polarized fermions caused by the scattering hypervolume $D_F$ in a periodic box.
Using this result we derive the thermodynamic properties, including the energy and the pressure, of the spin-polarized Fermi gas due to a nonzero $D_F$.

\subsection{Three fermions in a cubic box\label{subsec:three fermions in box}}
In this subsection, for the sake of simplicity we assume that the fermions have vanishing or negligible two-body interactions
but have a nonzero three-body scattering hypervolume $D_F$, and the $111$ expansion for the zero-energy three-body wave function in \Eq{111} is simplified as
\beq\label{Psi1Msimplified}
\Psi_1^M= Q_1^M(\vect s\times\vect R)\left(1-\frac{9\sqrt{3}D_F}{4\pi^3B^8}\right)+O(B^{-7}).
\eeq
For purposes of calculating the energy shifts due to a nonzero $D_F$, we can replace the true interaction potential $V(s_1)+V(s_2)+V(s_3)+V_{123}(s_1,s_2,s_3)$, which in general has a complicated dependence on the interparticle distances, by a three-body pseudopotential $V_{ps}$.
We use the following pseudopotential:
\begin{equation}\label{Vps}
    V_{ps}=\frac{\hbar^2D_F}{6M_F}\Big\{\Big[\nabla_{\vect{s}}^2\nabla_{\vect{R}}^2-(\nabla_{\vect{s}}\cdot\nabla_{\vect{R}})^2\Big]\delta(\vect{s})\delta(\vect{R})\Big\}\Lambda,
\end{equation}
where $\Lambda$ is a projection operator which, when acting on the $O(B^{-6})$ term in the three-body wave function, yields zero.
The operator $\Lambda$ is an analog of the operator $\frac{\partial}{\partial r}r$ in the two-body pseudopotential for $s$-wave two-body collisions
in Refs.~\cite{huang1957quantum,lee1957eigenvalues}.
One can check the pseudopotential in \Eq{Vps} is symmetric under the interchange of the three fermions. The coefficient on the right hand side of \Eq{Vps}
has been chosen such that
\begin{equation}\label{3bodySchrodingerVps}
\Big[-\frac{\hbar^2}{2M_F}(\nabla_1^2+\nabla_2^2+\nabla_3^2)+V_{ps}\Big]\Psi_1^M\simeq0.
\end{equation}

We now consider three fermions in a large periodic cubic box with volume $\Omega$. Their momenta are $\hbar\vect{k}_1$, $\hbar\vect{k}_2$ and $\hbar\vect{k}_3$ in the absence of interactions.
When we introduce interactions that give rise to a nonzero $D_F$, the energy eigenvalue of the three-body state is shifted
by the following amount at first order in the perturbation:
\begin{equation}
\mathcal{E}_{\vect{k}_1\vect{k}_2\vect{k}_3}=\int d^3 \vect{r}_1 d^3\vect{r}_2 d^3\vect{r}_3\, |\Psi_{\vect k_1\vect k_2\vect k_3}|^2 V_{ps},
\end{equation}
where $\Psi_{\vect k_1\vect k_2\vect k_3}$ is the normalized unperturbed wave function and it can be written as a Slater determinant:
\begin{equation}
\Psi_{\vect k_1\vect k_2\vect k_3}=\frac{1}{\sqrt{6}\Omega^{3/2}}
\left| \begin{matrix}
e^{\I \vect{k}_1\cdot\vect{r}_1}& e^{\I \vect{k}_1\cdot\vect{r}_2} & e^{\I \vect{k}_1\cdot\vect{r}_3}\\
e^{\I \vect{k}_2\cdot\vect{r}_1}& e^{\I \vect{k}_2\cdot\vect{r}_2} &e^{\I \vect{k}_2\cdot\vect{r}_3}  \\ 
e^{\I \vect{k}_3\cdot\vect{r}_1}& e^{\I \vect{k}_3\cdot\vect{r}_2} & e^{\I \vect{k}_3\cdot\vect{r}_3}
\end{matrix}\right|  .\label{Psi-Slater}
\end{equation}
We get
\begin{equation}\label{E of 3 fermions}
    \mathcal{E}_{\vect{k}_1\vect{k}_2\vect{k}_3}=\frac{\hbar^2 D_F}{3M_F \Omega^2}\left(\vect{k}_1\times\vect{k}_2+\vect{k}_2\times\vect{k}_3+\vect{k}_3\times\vect{k}_1\right)^2.
\end{equation}
Note that $|\vect{k}_1\times\vect{k}_2+\vect{k}_2\times\vect{k}_3+\vect{k}_3\times\vect{k}_1|$ is twice
the area of the $\vect k$-space triangle whose vertices are $\vect k_1,\vect k_2,\vect k_3$.

Note that \Eq{3bodySchrodingerVps} is only satisfied approximately.
In particular, if we take into account the $O(B^{-7})$ corrections in the asymptotic expansion of the wave function in \Eq{Psi1Msimplified}, \Eq{3bodySchrodingerVps} is violated.
So the three-body pseudopotential in \Eq{Vps} is only an approximate description of the effective three-body interaction.
However, we believe that this does \emph{not} affect our leading-order result for the energy shift in \Eq{E of 3 fermions}.
In Appendix~\ref{sec:3fermionenergy} we show another calculation, using Gauss's theorem, without resorting to the pseudopotential, that also gives rise to \Eq{E of 3 fermions}.

If there are two-body interactions, the shift of the energy of the three particles will also contain terms due to the two-body parameters including $a_p, r_p, a_f$ etc; nevertheless, the shift due to $D_F$ in \Eq{E of 3 fermions} is still valid. If $a_p\ne0$, the leading-order shift of the three-body energy due to $a_p$ is the sum of contributions from the three pairs of fermions.
To quickly derive this shift, we may write down the $p$-wave pseudopotential
\beq\label{Vps2body}
V_{ps}^{\text{2-body}}=\frac{6\pi\hbar^2a_p}{M_F}\big[\nabla^2\delta(\vect s)\big]\Lambda,
\eeq
where $s$ is the pairwise distance, and $\Lambda$ is the operator that annihilates the singular term that is $\propto s^{-2}$
in the two-body wave function $\phi^{(1,m)}(\vect s)$. The coefficient in this pseudopotential is chosen such that
the two-body wave function $\phi^{(1,m)}(\vect s)$ satisfies the Schr\"{o}dinger equation
$(-\frac{\hbar^2}{M_F}\nabla_\vect s^2+V_{ps}^\text{2-body})\phi^{(1,m)}(\vect s)=0$.
The pseudopotential that we write in \Eq{Vps2body} is similar to those given in Refs.~\cite{roth2001effective,kanjilal2004nondivergent,derevianko2005revised,stock2005generalized,pricoupenko2006modeling,idziaszek2006pseudopotential,reichenbach2006quasi,idziaszek2009analytical} but is of a simpler form.
Taking the expectation value of the pseudoptential in \Eq{Vps2body} in the unperturbed three-body state,
we derive the leading-order shift of the three-body energy due to a nonzero $a_p$:
\beq
\mathcal{E}^\text{2-body}_{\vect k_1\vect k_2\vect k_3}=\frac{6\pi\hbar^2a_p}{M_F\Omega}\big[(\vect k_1-\vect k_2)^2+(\vect k_2-\vect k_3)^2+(\vect k_3-\vect k_1)^2\big].
\eeq

\subsection{Energy shift of many fermions and thermodynamic consequences}

We generalize the energy shift in \Eq{E of 3 fermions} to $N$ fermions in the periodic volume $\Omega$. The number density of the fermions is $n=N/\Omega$.
We define the Fermi wave number $k_F=(6\pi^2 n)^{1/3}$, the Fermi energy $\epsilon_F=\hbar^2 k_F^2/2M_F$, and the Fermi temperature $T_F=\epsilon_F/k_B$. 
We assume that the density is low
such that the average interparticle distance $n^{-1/3}\gg r_e$.

\subsubsection{Adiabatic shifts of energy and pressure in the thermodynamic limit}
Starting from a many-body state at a finite temperature $T$, if we introduce a nonzero $D_F$ \emph{adiabatically}, the energy shift at first
order in $D_F$ is equal to the sum of the contributions from all the triples of fermions, namely
\begin{equation}
\Delta E=\frac{1}{6}\sum_{\vect{k}_1\vect{k}_2\vect{k}_3}\mathcal{E}_{\vect{k}_1\vect{k}_2\vect{k}_3}\, n_{\vect{k}_1}n_{\vect{k}_2}n_{\vect{k}_3},
\end{equation}
where $n_{\vect{k}}=(1+e^{\beta(\epsilon_{\vect{k}}-\mu)})^{-1}$ is the Fermi-Dirac distribution function, $\beta=1/k_B T$, $\epsilon_{\vect{k}}=\hbar^2 k^2/2M_F$ is the kinetic energy of a fermion with linear momentum $\hbar\vect k$, and $\mu$ is the chemical potential. The summation over $\vect{k}$ can be replaced by a continuous integral $\sum_{\vect{k}}=\Omega \int d^3k/(2\pi)^3$ in the thermodynamic limit. Carrying out the integral, we get
\begin{equation}
\Delta E(T)=\frac{N\hbar^2 D_F}{36\pi^4 M_F}k_F^{10}\cdot
 \left(\frac{9\pi}{64}\right) \widetilde{T}^5[f_{5/2}(e^{\beta\mu})]^2,
\end{equation}
where $\widetilde{T}=T/T_F$, and the function $f_{\nu}(z)$ is defined as
\begin{equation}
f_{\nu}(z)\equiv -\mathrm{Li}_{\nu}(-z)=z-\frac{z^2}{2^\nu}+\frac{z^3}{3^\nu}-\frac{z^4}{4^\nu}+\cdots.
\end{equation}
The chemical potential $\mu$ is determined by the number of fermions,
\begin{equation}
N=\Omega\int\frac{d^3k}{(2\pi)^3} \frac{1}{e^{\beta (\epsilon_{\vect{k}}-\mu)}+1},
\end{equation}
which is equivalent to
\beq
1=\frac{3\sqrt\pi}{4}\widetilde{T}^{3/2}f_{3/2}(e^{\widetilde{\mu}/\widetilde{T}}),
\eeq
where $\widetilde{\mu}=\mu/\epsilon_F$.

In the low temperature limit, $T\ll T_F$, 
\begin{equation}
\Delta E(T)=\frac{N\hbar^2 D_F}{36\pi^4 M_F}k_F^{10}\cdot
\Big[\frac{1}{25}+\frac{\pi^2}{30}\widetilde{T}^2+O(\widetilde{T}^4)\Big].
\end{equation}
In particular, at absolute zero temperature, 
\begin{equation}
\Delta E(0)=\frac{N\hbar^2 D_F}{900\pi^4 M_F}k_F^{10},
\end{equation}
and the ground state energy of the Fermi gas is
\begin{equation}
E=\frac{3}{5}\epsilon_F N \left(1+\frac{D_F k_F^8}{270\pi^4}+\cdots\right).
\end{equation}
If there are two-body interactions, the total ground state energy should contain terms that depend on the two-body parameters such as $a_p,r_p,a_f$, but the term due to $D_F$ remains the same as in the above formula to leading order in $D_F$.

In an intermediate temperature regime, $T_F\ll T\ll T_e$, 
\begin{equation}
\Delta E(T)=\frac{N\hbar^2 D_F}{36\pi^4 M_F}k_F^{10}
\bigg[\frac{1}{4}\widetilde{T}^2+\frac{\sqrt{\widetilde{T}}}{6\sqrt{2\pi}}+O(\widetilde{T}^{-1})\bigg].
\end{equation}
If $T$ is comparable to or higher than $T_e$, the de Broglie wave lengths of the fermions will be comparable to or shorter than the range $r_e$ of interparticle interaction potentials, and we can no longer use the effective parameter $D_F$ to describe the system.
See Fig. \ref{fig:energy} for $\Delta E$ as a function of the initial temperature.

The pressure of the spin-polarized Fermi gas changes by the following amount due to the adiabatic introduction of $D_F$:
\begin{equation}
\begin{split}
\Delta p(T)&=-\left(\frac{\partial \Delta E}{\partial \Omega}\right)_{S,N}
=\frac{10\Delta E}{3\Omega}\\
&=\frac{5n\hbar^2 D_F}{54\pi^4 M_F}k_F^{10}\cdot
\left(\frac{9\pi}{64}\right)\widetilde{T}^5  \left[f_{5/2}(e^{\beta\mu})\right]^2.\label{pressure adia}
\end{split}
\end{equation}
The subscripts $S,N$ in \Eq{pressure adia} mean that we keep the entropy $S$ and the particle number $N$ fixed when taking the partial derivative. See Fig. \ref{fig:pressure} for $\Delta p$ as a function of the initial temperature.
In particular, at zero temperature
\beq
\Delta p(0)=\frac{n\hbar^2 D_F}{270\pi^4 M_F}k_F^{10}.
\eeq

\subsubsection{Isothermal shifts of energy and pressure in the thermodynamic limit}
If the interaction is introduced adiabatically, the temperature will increase (if $D_F>0$) or decrease (if $D_F<0$). The change of temperature is
\begin{equation}
\Delta T=\left(\frac{\partial \Delta E}{\partial S}\right)_{N,\Omega}.
\end{equation}
Therefore, if we introduce $D_F$ isothermally, the energy shift $\Delta E'$ should be approximately
\begin{equation}
\Delta E'=\Delta E-C \Delta T=\left(1-T \frac{\partial}{\partial T}\right)\Delta E,
\end{equation}
where $C$ is the heat capacity of the noninteracting Fermi gas at constant volume.
In the low temperature limit, $T\ll T_F$,
\begin{equation}\label{DeltaE'lowT}
\Delta E'(T)=\frac{N\hbar^2 D_F}{36\pi^4 M_F}k_F^{10}\cdot
\Big[\frac{1}{25}-\frac{\pi^2}{30}\widetilde{T}^2+O(\widetilde{T}^4)\Big].
\end{equation}
In an intermediate temperature regime, $T_F\ll T\ll T_e$, 
\begin{equation}\label{DeltaE'highT}
\Delta E'(T)=\frac{N\hbar^2 D_F}{36\pi^4 M_F}k_F^{10}
\bigg[-\frac{1}{4}\widetilde{T}^2+\frac{\sqrt{\widetilde{T}}}{12\sqrt{2\pi}}+O(\widetilde{T}^{-1})\bigg].
\end{equation}
According to Eqs.~\eqref{DeltaE'lowT} and \eqref{DeltaE'highT}, $\Delta E'$ changes sign as we increase the temperature.
Therefore, there is a critical temperature $T_c$ at which $\Delta E'=0$. We find 
\begin{equation}
T_c \simeq 0.377 T_F.
\end{equation}

The pressure of the spin-polarized Fermi gas changes by the following amount due to the isothermal introduction of $D_F$:
\begin{equation}
\Delta p'=\Delta p-\frac{2 C\Delta T}{3\Omega}=
\left(1-\frac{1}{5}T \frac{\partial}{\partial T}\right)\Delta p.
\end{equation}
In the low temperature limit, $T\ll T_F$,
\begin{equation}
\Delta p'=\frac{5n\hbar^2 D_F}{54\pi^4 M_F}k_F^{10}\Big[\frac{1}{25}+\frac{\pi^2}{50}\widetilde{T}^2+O(\widetilde{T}^4)\Big].
\end{equation}
In an intermediate temperature regime, $T_F\ll T\ll T_e$,
\begin{equation}
\Delta p'=\frac{5n\hbar^2 D_F}{54\pi^4 M_F}k_F^{10}\bigg[\frac{3}{20}\widetilde{T}^2+\frac{3\sqrt{\widetilde{T}}}{20\sqrt{2\pi}}+O(\widetilde{T}^{-1})\bigg].
\end{equation}

\begin{figure}[htb]
\includegraphics[width=0.5\textwidth]{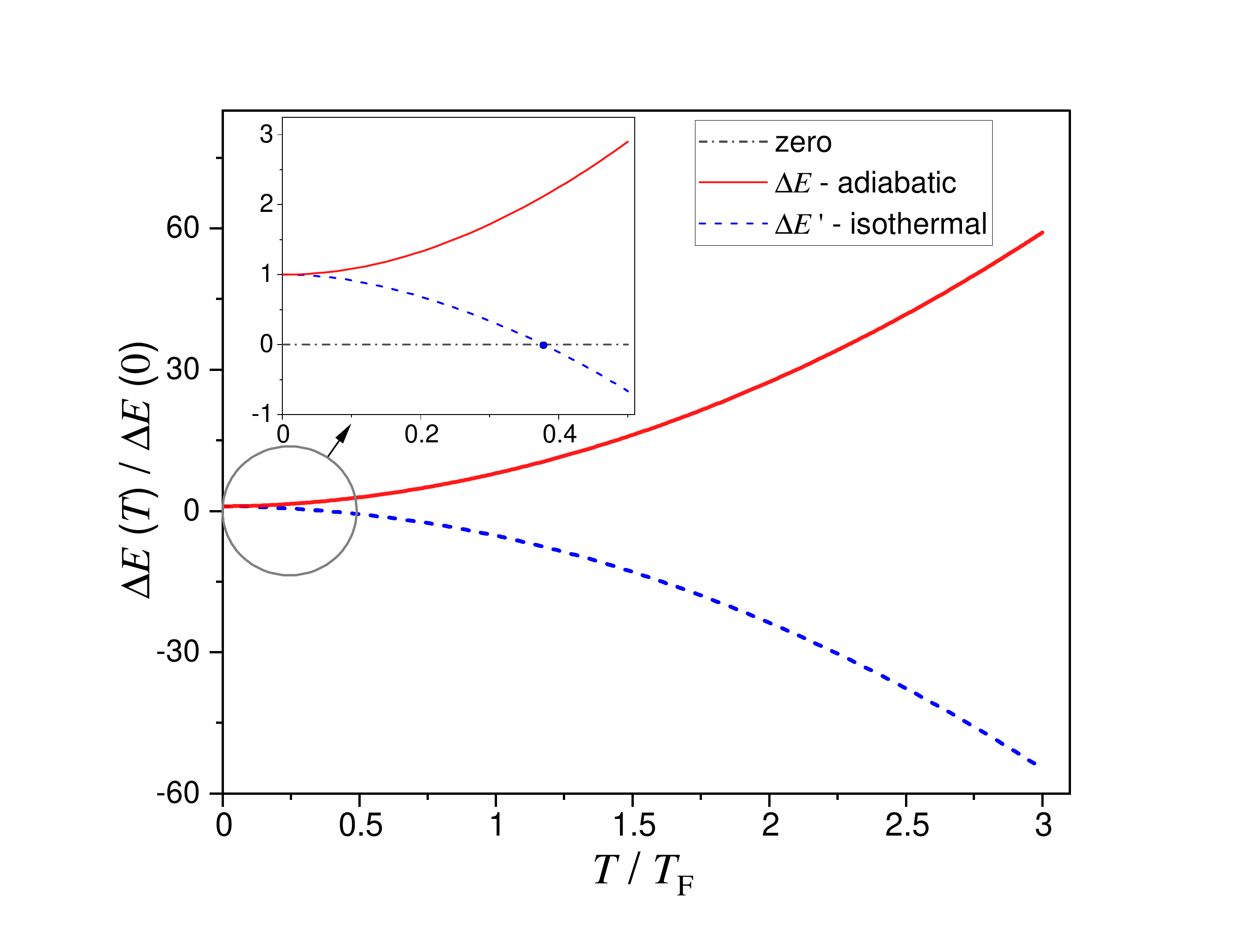}
\caption{\label{fig:energy}The energy shift caused by the adiabatic (red line) or isothermal (blue dashed line) introduction of $D_F$ versus the temperature $T$. At $T\simeq0.377 T_F$, the isothermal energy shift $\Delta E'$ changes sign.}
\end{figure}

\begin{figure}[htb]
\includegraphics[width=0.5\textwidth]{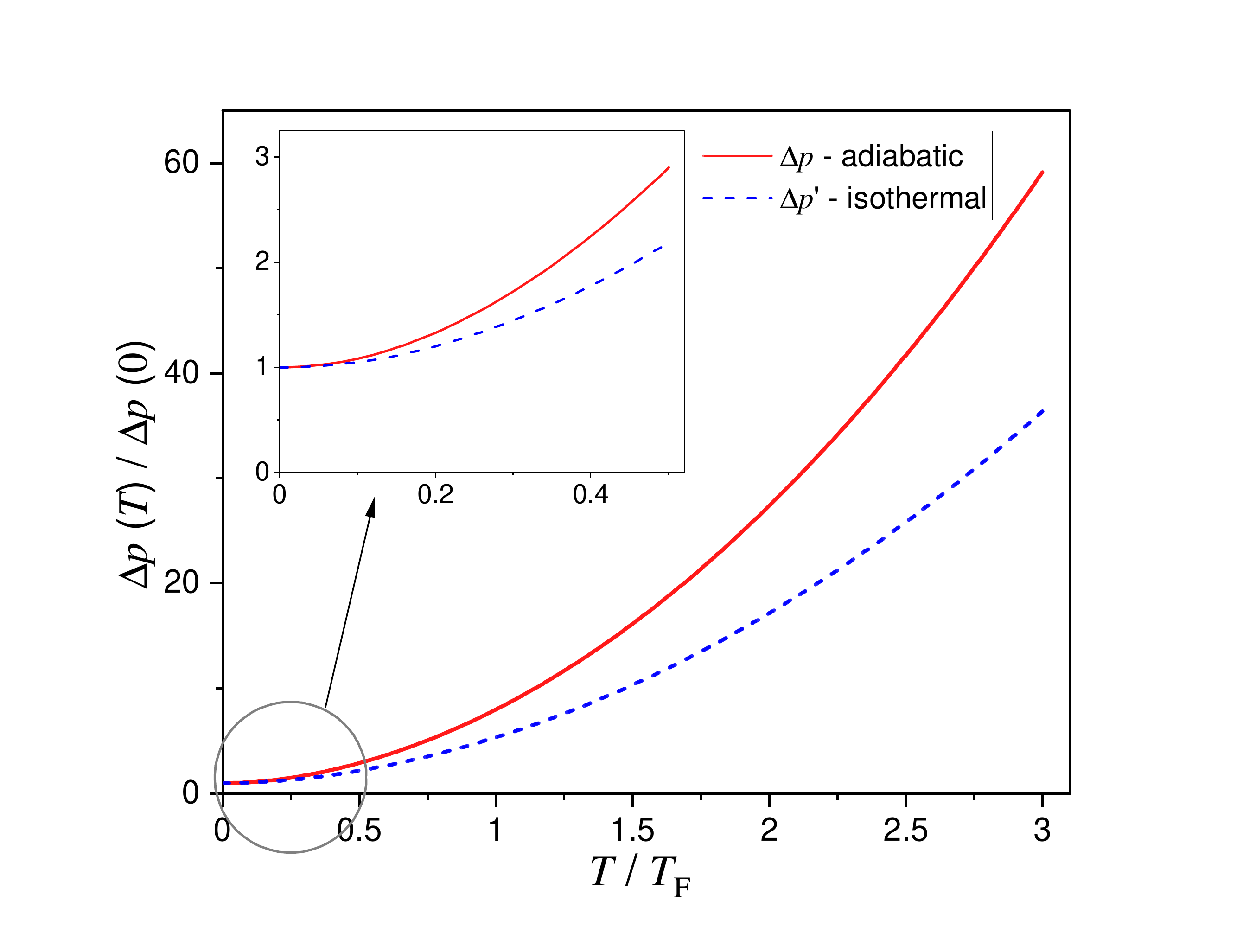}
\caption{\label{fig:pressure}The change of pressure caused by the adiabatic (red line) or isothermal (blue dashed line) introduction of $D_F$ versus the temperature $T$.}
\end{figure}

The energy shift and the pressure change as functions of temperature are shown in Fig. \ref{fig:energy} and Fig. \ref{fig:pressure} respectively.

\section{The Three-body recombination rate}\label{sec:recombination}
If the collision of the three particles is purely elastic, $D_F$ is a real number.
But if the two-body interactions support bound states, then the three-body collisions are usually not purely elastic, three-body recombination \cite{moerdijk1996decay,fedichev1996three,esry1999recombination,nielsen1999low,bedaque2000three,braaten2001three,HammerRevModPhys.85.197} will occur, and $D_F$ will acquire some negative imaginary part. The three-body recombination rate constant is proportional to the imaginary part of $D_F$ \cite{zhu2017three,braaten2006universality}.

Within a short time $\Delta t$, the probability that no recombination occurs is $\mathrm{exp}(-2|\mathrm{Im} {E}|\Delta t/\hbar)\simeq 1-2|\mathrm{Im} {E}|\Delta t/\hbar$. Then the probability that one recombination event occurs is $2|\mathrm{Im} {E}|\Delta t/\hbar$.
Since each recombination event causes the loss of three low-energy atoms, the change of the number of remaining low-energy atoms in the short time $\D t$ is
\begin{equation}
\D N=-\frac{1}{6}\sum_{\vect{k}_1\vect{k}_2\vect{k}_3}3\cdot\frac{2\D t}{\hbar}|\mathrm{Im}\mathcal{E}_{\vect{k}_1\vect{k}_2\vect{k}_3}| n_{\vect{k}_1}n_{\vect{k}_2}n_{\vect{k}_3}.
\end{equation}
This leads to
\begin{equation}
    \frac{\D n}{ \D t}=-L_3 n^{3},
\end{equation}
and the three-body recombination rate constant $L_3$ is
\begin{equation}
L_3=6 (6\pi^2)^{4/3} \left(\frac{9\pi}{64}\right)\widetilde{T}^5 [f_{5/2}(e^{\beta\mu})]^2 \frac{\hbar|\mathrm{Im} D_F|}{M_F}n^{4/3}.
\end{equation}
$L_3$ now in general depends on the density $n$ and the temperature $T$.

In the low temperature limit, $T\ll T_F$,
\begin{equation}\label{L3lowT}
L_3\simeq \frac{6 (6\pi^2)^{4/3}}{25}\left(1+\frac{5\pi^2}{6}\widetilde{T}^2\right)\frac{\hbar}{M_F}|\mathrm{Im} D_F| n^{4/3}.
\end{equation}
In this limit, $\D n/\D t$ is proportional to $n^{\alpha}$ where $\alpha=13/3=4.333\cdots$. In particular, at absolute zero temperature,
\begin{equation}
L_3=\frac{6}{25} \frac{\hbar|\mathrm{Im} D_F|}{M_F} k_F^4.
\end{equation}

In an intermediate temperature regime, $T_F\ll T\ll T_e$, we find that
\begin{equation}\label{L3highT}
L_3= 6\frac{M_F}{\hbar^3}|\mathrm{Im} D_F| \left(k_B T\right)^2,
\end{equation}
and $L_3$ is approximately independent of $n$, and is proportioanl to $T^2$.
It was predicted that $L_3\propto T^2$ according to the Wigner threshold law \cite{PhysRevA.65.010705}.
Our \Eq{L3highT} is consistent with this prediction. In Refs.~\cite{PhysRevLett.120.133401,top2020spinpolarized} the quadratic dependence of $L_3$ on the temperature was experimentally confirmed for ultracold $^6\mathrm{Li}$ atoms in the $|F=1/2,m_F=1/2\rangle$ state.
In Ref.~\cite{top2020spinpolarized} it was reported that $L_3=(3.55\pm0.22)\times10^{-23}\times T^2\mathrm{cm}^6/\mathrm{s}$ ($T$ in units of Kelvin) at $T\sim100-300\mu \mathrm{K}$ for $^6\mathrm{Li}$ atoms in a magnetic field of 1G (far away from the $p$-wave Feshbach resonance located at $159\,\mathrm{G}$). 
Using this result in Ref.~\cite{top2020spinpolarized} we find that the order of magnitude of $\mathrm{Im} D_F$
for these fermionic atoms in such magnetic field is about $-(125a_0)^8$, where $a_0$ is the Bohr radius.

In the intermediate temperature regime, $T_F\ll T\ll T_e$, the characteristic thermal de Broglie wave length $\lambda$ of the fermions is
shorter than the average interfermionic distance. Hence the orbital angular momentum quantum number for three colliding fermions may easily
exceed 1. One may wonder why the three-body recombination rate is still dominated by the parameter $D_F$ which refers to the $L=1$ collisions
only. We can resolve this paradox by noting that in this temperature regime $\lambda$ is still much
larger than the range $r_e$ of the interaction potentials. In the many-body wave function, when three fermions come to a spatial region
whose size is much smaller than the average interfermionic distance but much larger than $r_e$, the many-body wave function is approximately factorized as a 3-fermion Slater determinant [analogous to the one shown in \Eq{Psi-Slater}]
times a function that depends on the positions of the other fermions and the position of the center of mass of the three nearby fermions,
and when the three fermions come to distances smaller than $\lambda$ this 3-fermion Slater determinant may be further approximated by \Eq{C3} and \Eq{Phiqp} which in fact correspond to a collision with orbital angular momentum quantum number $L=1$. When the three fermions come to even smaller distances, the three-body wave function acquires
correction terms due to the interactions, and can be approximately described by the 111 expansion we have derived.
Therefore, the rate of three-body recombination events, which can occur only if three fermions come within the range of interaction,
is dominated by the parameter $D_F$ we have defined.

Ref.~\cite{castin2008threepolarized} considered spin-polarized fermionic atoms near a $p$-wave Feshbach resonance.
It was shown that when the two-body scattering volume $a_p$ is positive and large, such that there is a shallow two-body $p$-wave
bound state, the three-body recombination rate constant diverges as $a_p^{5/2}$ \cite{castin2008threepolarized}:
\beq
L_3\simeq\frac{9\hbar}{25M_F}(48\pi)^2\bigg(\frac{a_p^5}{-3r_p/2}\bigg)^{1/2}k_F^4
\eeq
at $T\ll T_F$.
Comparing this result with \Eq{L3lowT} we infer that near such a $p$-wave Feshbach resonance, on the $a_p>0$ side,
\beq
\im D_F\simeq-\frac{3}{2}(48\pi)^2\bigg(\frac{a_p^5}{-3r_p/2}\bigg)^{1/2}.
\eeq
Note that the $\alpha_\text{res}$ in Ref.~\cite{castin2008threepolarized} is equal to $(-r_p/2)$ in our paper,
and is positive.

\section{Summary and Discussion}
We have defined the three-body scattering hypervolume $D_F$ for identical spin-polarized fermions by considering the collision of three such fermions
at zero energy. We solved the three-body Schr\"{o}dinger equation asymptotically at large interparticle distances and expanded the three-body wave function in powers of $1/B$ when the pairwise distances go to infinity simultaneously (here $B$ is the three-body hyperradius), and expanded the same wave function in powers of $1/R$ when two fermions are held at a fixed distance and the third fermion is far away from the two (here $R$ is the distance between this third fermion and the center of mass of the other two fermions). In the expansion in powers of $1/B$, the parameter $D_F$ first appears at the order $1/B^6$. In the expansion in powers of $1/R$, the parameter $D_F$ first appears at the order $1/R^7$. For any given microscopic interaction potentials, one can solve the three-body Schr\"{o}dinger equation numerically and match the solution to these expansions to compute $D_F$.

The three-body scattering hypervolume we have defined in this paper plays a fundamental role in three-body, four-body, \dots, and many-body physics for 
ultracold Fermi gases. Although usually the two-body $p$-wave scattering volume is the dominant parameter for the effective interactions in spin-polarized ultracold Fermi gases, and $D_F$ serves as a small correction, $D_F$ may become the dominant parameter
if $a_p$ happens to be zero or tiny or is tuned to zero, or if the system is near a three-body resonance close to zero energy. If the system is near a three-body resonance close to zero energy, $D_F$ may be anomalously large.

In the second part of this paper 
we computed the energy shift of three fermions in a large periodic volume due to $D_F$. From this result we computed the energy shift of many fermions in the thermodynamic limit due to $D_F$. We also computed the shift of pressure due to $D_F$. The energy shift and the pressure shift are related. We computed the two shifts in two scenarios as functions of temperatures: either introducing the three-body parameter adiabatically or introducing it isothermally. For isothermal introduction of $D_F$, we found that the shift of energy changes sign at temperature $T\simeq 0.377 T_F$, where $T_F$ is the Fermi temperature. The energy and pressure shifts could be experimentally detected in trapped ultracold atomic Fermi gases. In particular, $D_F$ will cause a small change of the atomic cloud size and small changes of the collective mode frequencies.

If the two-body interactions are sufficiently attractive such that there are two-body bound states, $D_F$ will acquire some negative imaginary part related to the three-body recombination processes. We computed the three-body recombination rate constant $L_3$ in terms of $\im D_F$ as functions of temperature. In particular, we found that at low temperatures ($T\ll T_F$) $L_3\propto n^{4/3}$.
These results could be verified in future experiments concerning ultracold atomic Fermi gases.

\begin{acknowledgments}
This work was supported by the National Key R\&D Program of China (Grants No.~2019YFA0308403).
\end{acknowledgments}

\appendix
\section{Derivation of the 111 expansion and the 21 expansion for $L=1$, $M=0$}\label{Derivation}
We consider the collision of three identical spin-polarized fermions with orbital angular momentum $L=1$.
The magnetic quantum number $M$ can be $-1$, 0, and 1.
Here we first derive the 111 expansion and the 21 expansion for $M=0$.

We expand the three-body wave function in two forms:
\begin{subequations}
\begin{align}
    &\Psi=\sum_{p=-2}^{\infty} \mathcal{T}^{(-p)}(\vect{r}_1,\vect{r}_2,\vect{r}_3),\label{111-form}\\
    &\Psi=\sum_{q=-1}^{\infty}\mathcal{S}^{(-q)}(\vect{R},\vect{s}),\label{21-form}
\end{align}
\end{subequations}
where $\mathcal{T}^{(-p)}$ scales as $B^{-p}$, $\mathcal{S}^{(-q)}$ scales as $R^{-q}$. The hyperradius $B$ and the vectors $\vect{R}$ and $\vect{s}$ are already defined in the main text. 

If $s\ll R$, we can further expand $\mathcal{T}^{(-p)}$ as
\beq
\mathcal{T}^{(-p)}=\sum_{i} t^{(i,-p-i)},\label{T-p}
\eeq
where $t^{(i,j)}$ scales as $R^i s^j$.
If $s\gg r_e$, we can expand $\mathcal{S}^{(-q)}$ as
\beq
S^{(-q)}=\sum_{j} t^{(-q,j)}.
\eeq
Because the three-body wave function $\Psi_1^0$ may be expanded as $\sum_p \mathcal{T}^{(-p)}$ at $B\to\infty$,
and may also be expanded as $\sum_q \mathcal{S}^{(-q)}$ at $R\to\infty$, the $t^{(i,j)}$ in the above two expansions
should be the same. In fact the wave function has a double expansion $\Psi=\sum_{i,j}t^{(i,j)}$ in the region $r_e\ll s\ll R$.

\paragraph*{\textbf{Step 1.}}
We start from the leading-order term in the 111 expansion:
\beq
\mathcal{T}^{(2)}=Q_1^{0}(\vect{s}\times\vect{R})
=s_x R_y-s_y R_x=t^{(1,1)},\label{t11}
\end{equation}
and this indicates that $\mathcal{S}^{(1)}$ is nonzero, but $\mathcal{S}^{(2)}$, $\mathcal{S}^{(3)}$, $\mathcal{S}^{(4)}$, \dots are zero.
Consequently
\beq
t^{(i,j)}=0,~~\text{if}~i\ge2.
\eeq
Expanding $\mathcal{T}^{(2)}$ at $s\ll R$, we find that
\beq
t^{(0,2)}=t^{(-1,3)}=t^{(-2,4)}=t^{(-3,5)}=\cdots=0.
\eeq
Since $\mathcal{T}^{(3)}$, $\mathcal{T}^{(4)}$, $\mathcal{T}^{(5)}$, \dots are zero, we have
\beq
t^{(i,j)}=0,~~\text{if}~i+j\ge3.
\eeq

\paragraph*{\textbf{Step 2.}}
At $s\gg r_e$ we expand $\mathcal{S}^{(1)}$ as
\begin{equation}\label{S1expand}
\mathcal{S}^{(1)}=t^{(1,1)}+\sum_{j\le0}t^{(1,j)}.
\end{equation}
$\mathcal{S}^{(1)}$ also satisfies
\begin{equation}
\widetilde{H} \mathcal{S}^{(1)}=0,
\end{equation}
where $\widetilde{H}$ is proportional to the two-body Hamiltonian, and has been defined in the main text. Therefore, we have 
\begin{equation}
\mathcal{S}^{(1)}=R\sum_{l,m} c_m \phi ^{(l,m)}(\vect{s}). \label{S1}
\end{equation}
Here $l$ must be equal to 1, because $\phi ^{(l,m)}$ contributes a term proportional to $s^l$, and thus $S^{(1)}$ contains a term scaling as $R^1 s^l$. On the other hand, the leading order term on the right hand side of \Eq{S1expand} is $t^{(1,1)}$ which scales as $R^1s^1$. 

Expanding \Eq{S1} at $s\gg r_e$ to the order $s^1$, and using \Eq{two-body-phi}, we get
\begin{equation}
t^{(1,1)}=R\sum_m c_m \frac{s}{3} \sqrt{\frac{4\pi}{3}}Y_1^m (\hat{\vect{s}}).
\end{equation}
Comparing this result with \Eq{t11}, we find
\begin{subequations}
\begin{align}
&c_{-1}=\frac{-3 \I}{\sqrt{2} R} (R_x+\I R_y),\\
&c_{0}=0,\\
&c_{1}=\frac{-3 \I}{\sqrt{2} R} (R_x-\I R_y).
\end{align}
\end{subequations}
Therefore,
\begin{equation}
\mathcal{S}^{(1)}=\frac{-3 \I}{\sqrt{2}} (R_x+\I R_y)\phi ^{(1,-1)}(\vect s)+\frac{-3 \I}{\sqrt{2}} (R_x-\I R_y)\phi ^{(1,1)}(\vect s).
\end{equation}
The above result can be expressed in terms of the Clebsch-Gordan coefficients as
\begin{equation}
\mathcal{S}^{(1)}=6\I \sqrt{\frac{2\pi}{3}} R
\sum_m C_{1,-m;1,m}^{1,0}Y_1^{-m}(\hat{\vect{R}})\phi^{(1,m)}(\vect s).
\end{equation}
Expanding $\mathcal{S}^{(1)}$ at $s>r_e$, we get
\begin{subequations}
\begin{align}
&t^{(1,0)}=0,\\
&t^{(1,-1)}=0,\\
&t^{(1,-2)}=-\frac{3a_p}{s^3}\left(s_x R_y-s_y R_x\right),\label{t1,-2}\\
&t^{(1,j)}=0,~~j\le-3.
\end{align}
\end{subequations}

\paragraph*{\textbf{Step 3.}}
At $s\ll R$ we expand $\mathcal{T}^{(1)}$ as
\beq
\mathcal{T}^{(1)}=t^{(1,0)}+t^{(0,1)}+t^{(-1,2)}+\cdots=t^{(0,1)}+t^{(-1,2)}+\cdots.
\eeq
So $\mathcal T^{(1)}$ goes to zero at $s\to0$. So \Eq{free Schrodinger} may be written as $(\nabla_1^2+\nabla_2^2+\nabla_3^2)\mathcal T^{(1)}=0$ for $p=-1$,
and $\mathcal T^{(1)}$ should satisfy this partial differential equation even at $s_i=0$. Thus $\mathcal T^{(1)}$ must be a harmonic polynomial.
But we do \emph{not} have any nontrivial harmonic polynomial of degree 1 that also satisfies the fermionic antisymmetry.
We are therefore forced to take
\beq
\mathcal{T}^{(1)}=0.
\eeq
So
\beq
t^{(i,j)}=0,~~\text{if}~i+j=1.
\eeq

\paragraph*{\textbf{Step 4.}}
At $s\gg r_e$ we expand $\mathcal{S}^{(0)}$ as
\beq
\mathcal{S}^{(0)}=t^{(0,2)}+t^{(0,1)}+O(s^0)=O(s^0).
\eeq
Combining this with the equation $\widetilde{H}\mathcal{S}^{(0)}=0$, we get
\beq
\mathcal{S}^{(0)}=0.
\eeq
So
\beq
t^{(0,j)}=0.
\eeq

\paragraph*{\textbf{Step 5.}}
At $s\ll R$ we expand $\mathcal{T}^{(0)}$ as
\begin{align}
\mathcal{T}^{(0)}&=t^{(1,-1)}+t^{(0,0)}+t^{(-1,1)}+t^{(-2,2)}+\cdots\nonumber\\
&=t^{(-1,1)}+t^{(-2,2)}+\cdots.
\end{align}
So $\mathcal{T}^{(0)}$ goes to zero at $s\to0$. So \Eq{free Schrodinger} may be written as $(\nabla_1^2+\nabla_2^2+\nabla_3^2)\mathcal T^{(0)}=0$ for $p=0$,
and $\mathcal T^{(0)}$ should satisfy this partial differential equation even at $s_i=0$. Thus $\mathcal T^{(0)}$ must be a harmonic polynomial.
But we do \emph{not} have any nontrivial harmonic polynomial of degree 0 that also satisfies the fermionic antisymmetry.
We are therefore forced to take
\beq
\mathcal{T}^{(0)}=0.
\eeq
So
\beq
t^{(i,j)}=0,~~\text{if}~i+j=0.
\eeq

\paragraph*{\textbf{Step 6.}}
At $s\gg r_e$ we expand $\mathcal{S}^{(-1)}$ as
\beq
\mathcal{S}^{(-1)}=t^{(-1,3)}+t^{(-1,2)}+t^{(-1,1)}+O(s^0)=O(s^0).
\eeq
Combining this with the equation
\beq
\widetilde{H}\mathcal{S}^{(-1)}=\frac34\nabla_{\vect R}^2\mathcal{S}^{(1)}=0,
\eeq
we get
\beq
\mathcal{S}^{(-1)}=0.
\eeq
So
\beq
t^{(-1,j)}=0.
\eeq

\paragraph*{\textbf{Step 7.}}
At $s\ll R$ we expand $\mathcal{T}^{(-1)}$ as
\beq
\mathcal{T}^{(-1)}=t^{(1,-2)}+O(s^{-1}).
\eeq
$t^{(1,-2)}$ is shown in \Eq{t1,-2}.
$\mathcal{T}^{(-1)}$ should satisfy the free Schr\"odinger equation outside of the interaction range, so $(-\nabla_s^2 -3\nabla_R^2/4)\mathcal{T}^{(-1)}$ should be equal to some Dirac delta functions that are nonzero at $s_i=0$ only.
$\mathcal{T}^{(-1)}$ should also be antisymmetric under the interchange of the fermions.
We have 
$$-\nabla_s ^2 t^{(1,-2)}=12\pi a_p [R_y \partial_x \delta(\vect{s})-R_x \partial_y \delta(\vect{s})],$$
so
\begin{equation}
\left(-\nabla_\vect{s}^2-\frac{3}{4}\nabla_{\vect{R}}^2 \right)\mathcal{T}_{\vect{s}_1}^{(-1)}=12\pi a_p [R_y \partial_x \delta(\vect{s})-R_x \partial_y \delta(\vect{s})],
\end{equation}
where $\mathcal{T}_{\vect{s}_1}^{(-1)}$ is one term of the full $\mathcal{T}^{(-1)}$. 
Solving the above equation, we get
\begin{equation}
\mathcal{T}_{\vect{s}_1}^{(-1)}=\frac{-3a_p}{s_1^3}\left(s_x R_y-s_y R_x\right).
\end{equation}
The full $\mathcal{T}^{(-1)}$ should also be antisymmetric under the interchange of the fermions, so
\begin{equation}
\mathcal{T}^{(-1)}=-3a_p \left(s_x R_y-s_y R_x\right) \bigg(\frac{1}{s_1^3}+\frac{1}{s_2^3}+\frac{1}{s_3^3}\bigg).
\end{equation}

If $s\ll R$, we expand $\mathcal{T}^{(-1)}$ as $\sum_{n+m=-1} t^{(n,m)}$, and get
\begin{subequations}
\begin{align}
&t^{(-2,1)}=-6a_p \left(s_x R_y-s_y R_x\right)\frac{1}{R^3},\label{t_-2_1}\\
&t^{(-3,2)}=0,\\
&t^{(-4,3)}=9a_p\left(s_x R_y-s_y R_x\right) \frac{(R^2-5R_s^2)s^2}{4R^7},\label{t_-4_3}\\
&t^{(-5,4)}=0,\\
&t^{(-6,5)}=-45a_p\left(s_x R_y-s_y R_x\right)\nonumber\\
&\quad\quad\quad\quad \cdot\frac{(R^4-14R^2R_s^2+21R_s^4)s^4}{64R^{11}}\label{t_-6_5},\\
&t^{(-7,6)}=0,\\
&\cdots \nonumber
\end{align}
\end{subequations}
where $R_s\equiv \vect R\cdot\hat{\vect s}$.

\paragraph*{\textbf{Step 8.}}
At $s\gg r_e$ we expand $\mathcal{S}^{(-2)}$ as
\begin{align}
\mathcal{S}^{(-2)}&=t^{(-2,4)}+t^{(-2,3)}+t^{(-2,2)}+t^{(-2,1)}+\sum_{j\le0}t^{(-2,j)}\nn\\
&=t^{(-2,1)}+\sum_{j\le0}t^{(-2,j)}.\label{S-2expand}
\end{align}
$\mathcal{S}^{(-2)}$ satisfies the equation
\begin{equation}
\widetilde{H} \mathcal{S}^{(-2)}=\frac34\nabla_\vect R^2\mathcal{S}^{(0)}=0.
\end{equation}
So we get
\begin{equation}\label{S-2decompose}
\mathcal{S}^{(-2)}=\frac{1}{R^2}\sum_{l,m} d_m \phi ^{(l,m)}(\vect{s}).
\end{equation}
Here $l$ must be equal to 1, in order to be compatible with \Eq{S-2expand}.
Expanding \Eq{S-2decompose} at $s\gg r_e$, we find that $t^{(-2,1)}$ should be equal to 
\begin{equation}
t^{(-2,1)}=\frac{1}{R^2}\sum_m d_m \frac{s}{3} \sqrt{\frac{4\pi}{3}}Y_1^m (\hat{\vect{s}}).
\end{equation}
Comparing this with \Eq{t_-2_1}, we find
\begin{subequations}
\begin{align}
d_{-1}&=\frac{9\sqrt{2}\,a_p\I}{R}(R_x+\I R_y),\\
d_0&=0,\\
d_{1}&=\frac{9\sqrt{2}\,a_p\I}{R}(R_x-\I R_y).
\end{align}
\end{subequations}
Substituting these results into \Eq{S-2decompose}, we get
\begin{align}
 \mathcal{S}^{(-2)}=\frac{9\sqrt{2}\,a_p \I}{R^3} \bigg[&(R_x+\I R_y)\phi ^{(1,-1)}(\vect s)\nn\\
& + (R_x-\I R_y)\phi ^{(1,1)}(\vect s)\bigg]. 
\end{align}
This can be re-expressed in terms of the Clebsch-Gordan coefficients as
\begin{equation}
\mathcal{S}^{(-2)}=-\frac{36\I a_p}{R^2} \sqrt{\frac{2\pi}{3}}
\sum_m C_{1,-m;1,m}^{1,0}Y_1^{-m}(\hat{\vect{R}})\phi^{(1,m)}(\vect s).
\end{equation}
Expanding $\mathcal{S}^{(-2)}$ at $s\gg r_e$, we find
\begin{subequations}
\begin{align}
&t^{(-2,0)}=0,\\
&t^{(-2,-1)}=0,\\
&t^{(-2,-2)}=\frac{18a_p^2}{R^3s^3}\left(s_x R_y-s_y R_x\right),\\
&t^{(-2,j)}=0,~~j\le-3.
\end{align}
\end{subequations}

\paragraph*{\textbf{Step 9.}}
At $s\ll R$ we expand $\mathcal{T}^{(-2)}$ as
\beq
\mathcal{T}^{(-2)}=t^{(1,-3)}+t^{(0,-2)}+O(s^{-1})=O(s^{-1}).
\eeq
So the solution to the equation $(\nabla_1^2+\nabla_2^2+\nabla_3^2)\mathcal{T}^{(-1)}=0$ that is compatible with the above expansion
is
\beq
\mathcal{T}^{(-2)}=0.
\eeq
So
\beq
t^{(i,j)}=0,~~\text{if}~i+j=-2.
\eeq

\paragraph*{\textbf{Step 10.}}
At $s\gg r_e$ we expand $\mathcal{S}^{(-3)}$ as
\beq
\mathcal{S}^{(-3)}=\sum_{j\le5}t^{(-3,j)}=\sum_{j\le0}t^{(-3,j)}.
\eeq
Combining this with the equation
\beq
\widetilde{H}\mathcal{S}^{(-3)}=\frac34\nabla_\vect R^2\mathcal{S}^{(-1)}=0,
\eeq
we get
\beq
\mathcal{S}^{(-3)}=0.
\eeq
So
\beq
t^{(-3,j)}=0.
\eeq

\paragraph*{\textbf{Step 11.}}
At $s\ll R$ we expand $\mathcal{T}^{(-3)}$ as
\beq
\mathcal{T}^{(-3)}=t^{(1,-4)}+t^{(0,-3)}+t^{(-1,-2)}+O(s^{-1})=O(s^{-1}).
\eeq
So the solution to the equation $(\nabla_1^2+\nabla_2^2+\nabla_3^2)\mathcal{T}^{(-3)}=0$ that is compatible with the above expansion is
\beq
\mathcal{T}^{(-3)}=0.
\eeq
So
\beq
t^{(i,j)}=0,~~\text{if}~i+j=-3.
\eeq

\paragraph*{\textbf{Step 12.}}
At $s\gg r_e$ we expand $\mathcal{S}^{(-4)}$ as
\beq
\mathcal{S}^{(-4)}=\sum_{j\le6}t^{(-4,j)}=t^{(-4,3)}+\sum_{j\le0}t^{(-4,j)}.
\eeq
Combining this with the equation
\beq
\widetilde{H}\mathcal{S}^{(-4)}=\frac34\nabla_\vect R^2\mathcal{S}^{(-2)}=0,
\eeq
we find
\beq
\mathcal{S}^{(-4)}=-\frac{630\sqrt{\pi}\,\I a_p}{R^4}\sum_m C_{3,-m;3,m}^{1,0}Y_3^{-m}(\hat{\vect{R}})\phi^{(3,m)}(\vect s).
\eeq
Expanding $\mathcal{S}^{(-4)}$ at $s\gg r_e$, we get
\begin{subequations}
\begin{align}
&t^{(-4,0)}=0,\\
&t^{(-4,-1)}=0,\\
&t^{(-4,-2)}=0.
\end{align}
\end{subequations}

\paragraph*{\textbf{Step 13.}}
At $s\ll R$ we expand $\mathcal{T}^{(-4)}$ as
\begin{align}
\mathcal{T}^{(-4)}&=t^{(1,-5)}+t^{(0,-4)}+t^{(-1,-3)}+t^{(-2,-2)}+O(s^{-1})\nn\\
&=\frac{18a_p^2}{R^3s^3}\left(s_x R_y-s_y R_x\right)+O(s^{-1}).
\end{align}
The solution to the equation $(\nabla_1^2+\nabla_2^2+\nabla_3^2)\mathcal{T}^{(-4)}=0$ that is compatible
with the above expansion is
\begin{equation}
\mathcal{T}^{(-4)}=\frac{36a_p^2}{\pi}\left(s_x R_y-s_y R_x\right)\sum_{i=1}^3 \frac{\theta_i-\frac{1}{4}\sin 4\theta_i}{R_i^3 s_i^3}.
\end{equation}


Expanding $\mathcal{T}^{(-4)}$ at $s\ll R$ as $\sum_{i+j=-4} t^{(i,j)}$, we get
\begin{subequations}
\begin{align}
&t^{(-3,-1)}=0,\\
&t^{(-4,0)}=0,\\
&t^{(-5,1)}=\frac{12a_p^2}{\pi} \left(s_x R_y-s_y R_x\right)\frac{8\pi-9\sqrt{3}}{R^6},\label{t_-5_1}\\
&t^{(-6,2)}=0,\\
&t^{(-7,3)}=\frac{9a_p^2}{5\pi}\left(s_x R_y-s_y R_x\right)\nonumber\\
&\quad\quad\quad \cdot\frac{[(297\sqrt{3}-200\pi)R^2+(640\pi-1080\sqrt{3})R_s^2]s^2}{R^{10}},\label{t_-7_3}\\
&\cdots \nonumber
\end{align}
\end{subequations}


\paragraph*{\textbf{Step 14.}}
At $s\gg r_e$ we expand $\mathcal{S}^{(-5)}$ as
\begin{align}
\mathcal{S}^{(-5)}&=\sum_{j\le 7}t^{(-5,j)}\nn\\
&=\frac{12a_p^2}{\pi} \left(s_x R_y-s_y R_x\right)\frac{8\pi-9\sqrt{3}}{R^6}+\sum_{j\le0}t^{(-5,j)}.
\end{align}
Combining this with the equation
\beq
\widetilde{H}\mathcal{S}^{(-5)}=\frac34\nabla_\vect R^2\mathcal{S}^{(-3)}=0,
\eeq
we get
\begin{equation}
\begin{split}
\mathcal{S}^{(-5)}=&\frac{72 (8-\frac{9\sqrt{3}}{\pi})\I a_p^2}{R^5}\sqrt{\frac{2\pi}{3}}\\
&\times\sum_m C_{1,-m;1,m}^{1,0}Y_1^{-m}(\hat{\vect{R}})\phi^{(1,m)}(\vect s).
\end{split}
\end{equation}
So
\begin{subequations}
\begin{align}
&t^{(-5,0)}=0,\\
&t^{(-5,-1)}=0.
\end{align}
\end{subequations}

\paragraph*{\textbf{Step 15.}}
At $s\ll R$ we expand $\mathcal{T}^{(-5)}$ as
\beq
\mathcal{T}^{(-5)}=\sum_{i\le 1}t^{(i,-5-i)}=O(s^{-1}).
\eeq
The solution to the equation $(\nabla_1^2+\nabla_2^2+\nabla_3^2)\mathcal{T}^{(-5)}=0$ that is compatible
with the above expansion is
\beq
\mathcal{T}^{(-5)}=0.
\eeq
So
\beq
t^{(i,j)}=0,~~\text{if}~i+j=-5.
\eeq


\paragraph*{\textbf{Step 16.}}
At $s\gg r_e$ we expand $\mathcal{S}^{(-6)}$ as
\begin{align}
\mathcal{S}^{(-6)}&=\sum_{j\le8}t^{(-6,j)}=t^{(-6,5)}+O(s^0).
\end{align}
Combining this with the equation
\beq
\widetilde{H}\mathcal{S}^{(-6)}=\frac34\nabla_\vect R^2\mathcal{S}^{(-4)}=0,
\eeq
we get
\begin{align}
\mathcal{S}^{(-6)}=&-\frac{31185\sqrt{10\pi}\,\I a_p}{4R^6}\nn\\
&\times\sum_m C_{5,-m;5,m}^{1,0}Y_5^{-m}(\hat{\vect{R}})\phi^{(5,m)}(\vect s).
\end{align}
So
\beq
t^{(-6,0)}=0.
\eeq

\paragraph*{\textbf{Step 17.}}
At $s\ll R$ we expand $\mathcal{T}^{(-6)}$ as
\beq
\mathcal{T}^{(-6)}=\sum_{i\le1}t^{(i,-6-i)}=O(s^{-1}).
\eeq
The solution to the equation $(\nabla_1^2+\nabla_2^2+\nabla_3^2)\mathcal{T}^{(-6)}=0$ (for $B>0$) that is compatible
with the above expansion is
\begin{equation}
\mathcal{T}^{(-6)}=-\frac{9\sqrt{3}D_F}{4\pi^3 B^8} \left( s_x R_y- s_y R_x\right).
\end{equation}
It satisfies
\begin{equation}
\begin{split}
&\left( -\nabla_{\vect{s}}^2-\frac{3}{4}\nabla_\vect{R}^2\right)\mathcal{T}^{(-6)}\\
&=-D_F \Big[\frac{\partial\delta(\vect{s})}{\partial s_x}\frac{\partial \delta (\vect{R})}{\partial R_y}-\frac{\partial \delta(\vect{s})}{\partial s_y}\frac{\partial \delta(\vect{R})}{\partial R_x}\Big].
\end{split}
\end{equation}
Expanding $\mathcal{T}^{(-6)}$ at $s\ll R$, we get
\begin{subequations}
\begin{align}
&t^{(-7,1)}=-\frac{9\sqrt{3}D_F}{4\pi^3 R^8} \left( s_x R_y- s_y R_x\right),\label{t_-7_1}\\
&\cdots \nonumber
\end{align}
\end{subequations}


\paragraph*{\textbf{Step 18.}}
At $s\gg r_e$ we expand $\mathcal{S}^{(-7)}$ as
\beq
\mathcal{S}^{(-7)}=\sum_{j\le9}t^{(-7,j)}=t^{(-7,3)}+t^{(-7,1)}+O(s^0).
\eeq
Combining this with the equation
\begin{align}
&\widetilde{H}\mathcal{S}^{(-7)}=\frac34\nabla_\vect R^2\mathcal{S}^{(-5)}\nn\\
&=\frac{324\sqrt{6\pi}(8-\frac{9\sqrt{3}}{\pi})\I a_p^2}{R^7}\sum_m C_{1,-m;1,m}^{1,0}Y_1^{-m}(\hat{\vect{R}})\phi^{(1,m)}(\vect s),
\end{align}
we get
\begin{equation}
\begin{split}
&\mathcal{S}^{(-7)}\\
&=\frac{324\sqrt{6\pi}(8-\frac{9\sqrt{3}}{\pi})a_p^2 \I}{R^7}\sum_m C_{1,-m;1,m}^{1,0}Y_1^{-m}(\hat{\vect{R}})f^{(1,m)}(\vect s)\\
&\quad-\frac{6\xi \I}{R^7}\sqrt{\frac{2\pi}{3}}\sum_m C_{1,-m;1,m}^{1,0}Y_1^{-m}(\hat{\vect{R}})\phi^{(1,m)}(\vect s)\\
&\quad+\frac{4032\sqrt{\pi}(16-\frac{27\sqrt{3}}{\pi})a_p^2\I}{R^7}\\
&\quad\quad\times\sum_m C_{3,-m;3,m}^{1,0}Y_3^{-m}(\hat{\vect{R}})\phi^{(3,m)}(\vect s),
\end{split}
\end{equation}
where $\xi$ is a parameter related to $D_F$, 
\begin{equation}
\xi=\frac{9\sqrt{3}D_F}{4\pi^3}-81a_p^3 r_p\left( 8-\frac{9\sqrt{3}}{\pi}\right) .
\end{equation}


We have thus derived the 111 expansion to the order $B^{-6}$ and the 21 expansion to the order $R^{-7}$.

\section{The 111 expansion and the 21 expansion for $L=1$, $M=\pm 1$\label{sec:M=pm1}}
For the magnetic quantum number $M=\pm 1$, one can start from the leading order term
\beq
\mathcal{T}^{(2)}=Q_1^{\pm 1}(\vect{s}\times\vect{R}),
\eeq
and do the same type of step-by-step derivation as in the last section.
The details of the derivation are similar.
For brevity we do not show the details.

We can also use the ladder operators $J_{\pm}$,
\begin{equation}
J_{\pm}=J_x \pm \I J_y,
\end{equation}
where $J_x$ and $J_y$ are the projections of the total orbital angular momentum operator in the $x$ and $y$ directions, respectively.
Applying the ladder operator $J_+$ or $J_-$ to the 111 expansion and the 21 expansion for $M=0$, 
we get the corresponding expansions for $M=1$ or $M=-1$.

\section{An alternative method for the derivation of the energy of three fermions in a large box\label{sec:3fermionenergy}}
The wave function of three free fermions with momenta $\hbar\vect{k}_1,\hbar\vect{k}_2,\hbar\vect{k}_3$ in a large periodic cubic box is
\begin{equation}\label{freePsi}
\Psi_{\vect k_1\vect k_2\vect k_3}=\frac{1}{\sqrt{6}\Omega^{3/2}}
\left| \begin{matrix}
e^{\I \vect{k}_1\cdot\vect{r}_1}& e^{\I \vect{k}_1\cdot\vect{r}_2} & e^{\I \vect{k}_1\cdot\vect{r}_3}\\
e^{\I \vect{k}_2\cdot\vect{r}_1}& e^{\I \vect{k}_2\cdot\vect{r}_2} &e^{\I \vect{k}_2\cdot\vect{r}_3}  \\ 
e^{\I \vect{k}_3\cdot\vect{r}_1}& e^{\I \vect{k}_3\cdot\vect{r}_2} & e^{\I \vect{k}_3\cdot\vect{r}_3}
\end{matrix}\right|  .
\end{equation}
We define the Jacobi momenta $\hbar\vect{q},\hbar\vect{p},\hbar\vect{k}_c$ such that
\begin{subequations}
\begin{align}
&\vect{k}_1=\frac{1}{3}\vect{k}_c+\frac{1}{2}\vect{q}+\vect{p},\\
&\vect{k}_2=\frac{1}{3}\vect{k}_c+\frac{1}{2}\vect{q}-\vect{p},\\
&\vect{k}_3=\frac{1}{3}\vect{k}_c-\vect{q}.
\end{align}
\end{subequations}
$\hbar\vect{k}_c$ is the total momentum of three fermions. We extract the motion of the center of mass $\vect{R}_c=(\vect r_1+\vect r_2+\vect r_3)/3$, 
\begin{equation}
\Psi_{\vect{k}_1\vect{k}_2\vect{k}_3}=
\frac{1}{\sqrt{\Omega}}
e^{\I \vect{k}_c\cdot\vect{R}_c}\Phi_{\vect{q},\vect{p}}.\label{C3}
\end{equation}
Suppose that the typical momentum of each fermion is $\sim2\pi\hbar/\lambda$.
For small hyperradii, $B\ll\lambda$, we Taylor expand $\Phi_{\vect q,\vect p}$ and get
\begin{equation}
\Phi_{\vect{q},\vect{p}}\simeq \frac{3}{\sqrt{6}\Omega}(\vect{p}\times\vect{q})\cdot(\vect{s}_3\times\vect{R}_3).\label{Phiqp}
\end{equation}
$\Phi_{\vect{q},\vect{p}}$ is the wave function of the relative motion of three free fermions. If we introduce a small three-body $D_F$ adiabatically, it is changed to
\begin{equation}\label{relativePhi}
\Phi_{\vect{q},\vect{p}}\simeq \frac{3}{\sqrt{6}\Omega}(\vect{p}\times\vect{q})\cdot(\vect{s}_3\times\vect{R}_3)\left(1-\frac{9\sqrt{3}D_F}{4\pi^3 B^8}\right)
\end{equation}
for $r_e\ll B\ll\lambda$, where $r_e$ is the range of interaction. The wave function satisfies the free Schr\"{o}dinger equation outside of the range of interaction,
\begin{equation}\label{SEinbox}
-\frac{\hbar^2}{M_F}\nabla_{\bm\rho}^2\Phi_{\vect{q},\vect{p}}=E\Phi_{\vect{q},\vect{p}},
\end{equation}
where $\bm\rho=(\vect{s},2\vect{R}/\sqrt{3})$ is a six dimensional vector, $E$ is the energy of the relative motion, and $B=\sqrt{3}\rho/2$.

For large box sizes, we may compute the energy $E$ approximately. We rewrite \Eq{SEinbox} as
\begin{subequations}
\begin{align}
&-\frac{\hbar^2}{M_F}\nabla_{\bm\rho}^2\Phi_{1}=E_1\Phi_{1},\label{e1}\\
&-\frac{\hbar^2}{M_F}\nabla_{\bm\rho}^2\Phi_{2}^{*}=E_2\Phi_{2}^{*},\label{e2}
\end{align}
\end{subequations}
for two different interactions that yield two different three-body scattering hypervolumes, $D_{F1}$ and $D_{F2}$ respectively.
Here we omit the subscript $\vect{q},\vect{p}$ for simplicity. Multiplying both sides of \Eq{e1} by $\Phi_2^*$,
multiplying both sides of \Eq{e2} by $\Phi_1$, subtracting the two resultant equations, and taking six-dimensional integral over $\bm\rho$ for $\rho>\rho_0$
(where $\rho_0$ is any length scale satisfying $r_e\ll\rho_0\ll\lambda$), we get
\begin{align}
&-\frac{\hbar^2}{M_F}\int_{\rho>\rho_0}\!\!\! d^6 \rho~ \nabla_{\bm\rho}\cdot(\Phi_2^*\nabla_{\bm\rho} \Phi_1-\Phi_1\nabla_{\bm\rho}\Phi_2^*)\nonumber\\
&=(E_1-E_2)\int_{\rho>\rho_0}\!\!\! d^6\rho~ \Phi_1 \Phi_2^*.\label{Gauss}
\end{align}
In the region $\rho>\rho_0$, $\Phi_{1}\simeq\Phi_2$, and the right hand side of \Eq{Gauss} is
\begin{equation}
(E_1-E_2)\frac{8}{3\sqrt{3}}\int_{\rho>\rho_0}\!\!\! d^3 \vect{s} ~d^3 \vect{R}~ |\Phi|^2
\simeq\frac{8}{3\sqrt{3}}(E_1-E_2)
\end{equation}
because the wave function for the relative motion is normalised, and the volume of the region $\rho<\rho_0$ is small and may be omitted in the integral. Applying Gauss's divergence theorem to the left hand side of \Eq{Gauss}, we get
\begin{equation}\label{intonS}
-\frac{\hbar^2}{M_F}\oint_{\rho=\rho_0}\!\!\! d\vect{S}\cdot (\Phi_2^*\nabla_{\bm\rho} \Phi_1-\Phi_1\nabla_{\bm\rho}\Phi_2^*)\simeq\frac{8}{3\sqrt{3}}(E_1-E_2),
\end{equation}
where $S$ is the surface of the hypersphere with radius $\rho=\rho_0$,
and $d\vect S$ points toward the center of the hypersphere.

To evaluate the integral on the surface, we parametrize the six coordinates $\bm\rho=(\rho^{(1)},\rho^{(2)},\rho^{(3)},\rho^{(4)},\rho^{(5)},\rho^{(6)})$ as
\begin{subequations}
\begin{align}
&\rho^{(1)}=\rho \cos \varphi_1,\\
&\rho^{(2)}=\rho \sin\varphi_1\cos \varphi_2,\\
&\rho^{(3)}=\rho \sin\varphi_1\sin\varphi_2\cos \varphi_3,\\
&\rho^{(4)}=\rho \sin\varphi_1\sin\varphi_2\sin \varphi_3\cos\varphi_4,\\
&\rho^{(5)}=\rho \sin\varphi_1\sin\varphi_2\sin \varphi_3\sin\varphi_4\cos\varphi_5,\\
&\rho^{(6)}=\rho \sin\varphi_1\sin\varphi_2\sin \varphi_3\sin\varphi_4\sin\varphi_5,
\end{align}
\end{subequations}
where $0\leq\varphi_1,\cdots\varphi_4\leq\pi$, and $0\leq\varphi_5<2\pi$. Here
$\vect{s}=(\rho^{(1)},\rho^{(2)},\rho^{(3)})$ and $\vect{R}=\frac{\sqrt{3}}{2} (\rho^{(4)},\rho^{(5)},\rho^{(6)})$. The surface element $d\vect{S}$ is
\begin{align}
d\vect{S}=&-\hat{\bm\rho} \rho^5 d\varphi_1 d\varphi_2 d\varphi_3 d\varphi_4 d\varphi_5 \nonumber\\
&\cdot\sin^4\varphi_1 \sin^3\varphi_2 \sin^2\varphi_3 \sin\varphi_4 .
\end{align}
The minus sign in the above equation means that the direction of $d\vect{S}$ is towards the origin.  
Using the coordinates $\rho,\varphi_1,\cdots\varphi_5$, we rewrite the wave function in \Eq{relativePhi}, and evaluate the integral on the hypersphere with radius $\rho=\rho_0$. We get
\begin{equation}
\begin{split}
&E_2-E_1=\frac{3\hbar^2}{M_F\Omega^2}(D_{F2}-D_{F1})(\vect{p}\times\vect{q})^2\\
&\simeq\frac{\hbar^2(D_{F2}-D_{F1})}{3M_F\Omega^2}\left(\vect{k}_1\times\vect{k}_2+\vect{k}_2\times\vect{k}_3+\vect{k}_3\times\vect{k}_1\right)^2.
\end{split}
\end{equation}
This result agrees with \Eq{E of 3 fermions}.

\bibliography{ref}

\providecommand{\noopsort}[1]{}\providecommand{\singleletter}[1]{#1}%
\begin{thebibliography}{33}%
\makeatletter
\providecommand \@ifxundefined [1]{%
 \@ifx{#1\undefined}
}%
\providecommand \@ifnum [1]{%
 \ifnum #1\expandafter \@firstoftwo
 \else \expandafter \@secondoftwo
 \fi
}%
\providecommand \@ifx [1]{%
 \ifx #1\expandafter \@firstoftwo
 \else \expandafter \@secondoftwo
 \fi
}%
\providecommand \natexlab [1]{#1}%
\providecommand \enquote  [1]{``#1''}%
\providecommand \bibnamefont  [1]{#1}%
\providecommand \bibfnamefont [1]{#1}%
\providecommand \citenamefont [1]{#1}%
\providecommand \href@noop [0]{\@secondoftwo}%
\providecommand \href [0]{\begingroup \@sanitize@url \@href}%
\providecommand \@href[1]{\@@startlink{#1}\@@href}%
\providecommand \@@href[1]{\endgroup#1\@@endlink}%
\providecommand \@sanitize@url [0]{\catcode `\\12\catcode `\$12\catcode
  `\&12\catcode `\#12\catcode `\^12\catcode `\_12\catcode `\%12\relax}%
\providecommand \@@startlink[1]{}%
\providecommand \@@endlink[0]{}%
\providecommand \url  [0]{\begingroup\@sanitize@url \@url }%
\providecommand \@url [1]{\endgroup\@href {#1}{\urlprefix }}%
\providecommand \urlprefix  [0]{URL }%
\providecommand \Eprint [0]{\href }%
\providecommand \doibase [0]{https://doi.org/}%
\providecommand \selectlanguage [0]{\@gobble}%
\providecommand \bibinfo  [0]{\@secondoftwo}%
\providecommand \bibfield  [0]{\@secondoftwo}%
\providecommand \translation [1]{[#1]}%
\providecommand \BibitemOpen [0]{}%
\providecommand \bibitemStop [0]{}%
\providecommand \bibitemNoStop [0]{.\EOS\space}%
\providecommand \EOS [0]{\spacefactor3000\relax}%
\providecommand \BibitemShut  [1]{\csname bibitem#1\endcsname}%
\let\auto@bib@innerbib\@empty
\bibitem [{\citenamefont {Tan}(2008)}]{tan2008three}%
  \BibitemOpen
  \bibfield  {author} {\bibinfo {author} {\bibfnamefont {S.}~\bibnamefont
  {Tan}},\ }\bibfield  {title} {\bibinfo {title} {Three-boson problem at low
  energy and implications for dilute bose-einstein condensates},\ }\href
  {https://doi.org/10.1103/PhysRevA.78.013636} {\bibfield  {journal} {\bibinfo
  {journal} {Phys. Rev. A}\ }\textbf {\bibinfo {volume} {78}},\ \bibinfo
  {pages} {013636} (\bibinfo {year} {2008})}\BibitemShut {NoStop}%
\bibitem [{\citenamefont {Moerdijk}\ \emph {et~al.}(1996)\citenamefont
  {Moerdijk}, \citenamefont {Boesten},\ and\ \citenamefont
  {Verhaar}}]{moerdijk1996decay}%
  \BibitemOpen
  \bibfield  {author} {\bibinfo {author} {\bibfnamefont {A.~J.}\ \bibnamefont
  {Moerdijk}}, \bibinfo {author} {\bibfnamefont {H.~M. J.~M.}\ \bibnamefont
  {Boesten}},\ and\ \bibinfo {author} {\bibfnamefont {B.~J.}\ \bibnamefont
  {Verhaar}},\ }\bibfield  {title} {\bibinfo {title} {Decay of trapped
  ultracold alkali atoms by recombination},\ }\href
  {https://doi.org/10.1103/PhysRevA.53.916} {\bibfield  {journal} {\bibinfo
  {journal} {Phys. Rev. A}\ }\textbf {\bibinfo {volume} {53}},\ \bibinfo
  {pages} {916} (\bibinfo {year} {1996})}\BibitemShut {NoStop}%
\bibitem [{\citenamefont {Fedichev}\ \emph {et~al.}(1996)\citenamefont
  {Fedichev}, \citenamefont {Reynolds},\ and\ \citenamefont
  {Shlyapnikov}}]{fedichev1996three}%
  \BibitemOpen
  \bibfield  {author} {\bibinfo {author} {\bibfnamefont {P.~O.}\ \bibnamefont
  {Fedichev}}, \bibinfo {author} {\bibfnamefont {M.~W.}\ \bibnamefont
  {Reynolds}},\ and\ \bibinfo {author} {\bibfnamefont {G.~V.}\ \bibnamefont
  {Shlyapnikov}},\ }\bibfield  {title} {\bibinfo {title} {Three-body
  recombination of ultracold atoms to a weakly bound $\mathit{s}$ level},\
  }\href {https://doi.org/10.1103/PhysRevLett.77.2921} {\bibfield  {journal}
  {\bibinfo  {journal} {Phys. Rev. Lett.}\ }\textbf {\bibinfo {volume} {77}},\
  \bibinfo {pages} {2921} (\bibinfo {year} {1996})}\BibitemShut {NoStop}%
\bibitem [{\citenamefont {Esry}\ \emph {et~al.}(1999)\citenamefont {Esry},
  \citenamefont {Greene},\ and\ \citenamefont {Burke}}]{esry1999recombination}%
  \BibitemOpen
  \bibfield  {author} {\bibinfo {author} {\bibfnamefont {B.~D.}\ \bibnamefont
  {Esry}}, \bibinfo {author} {\bibfnamefont {C.~H.}\ \bibnamefont {Greene}},\
  and\ \bibinfo {author} {\bibfnamefont {J.~P.}\ \bibnamefont {Burke}},\
  }\bibfield  {title} {\bibinfo {title} {Recombination of three atoms in the
  ultracold limit},\ }\href {https://doi.org/10.1103/PhysRevLett.83.1751}
  {\bibfield  {journal} {\bibinfo  {journal} {Phys. Rev. Lett.}\ }\textbf
  {\bibinfo {volume} {83}},\ \bibinfo {pages} {1751} (\bibinfo {year}
  {1999})}\BibitemShut {NoStop}%
\bibitem [{\citenamefont {Nielsen}\ and\ \citenamefont
  {Macek}(1999)}]{nielsen1999low}%
  \BibitemOpen
  \bibfield  {author} {\bibinfo {author} {\bibfnamefont {E.}~\bibnamefont
  {Nielsen}}\ and\ \bibinfo {author} {\bibfnamefont {J.~H.}\ \bibnamefont
  {Macek}},\ }\bibfield  {title} {\bibinfo {title} {Low-energy recombination of
  identical bosons by three-body collisions},\ }\href
  {https://doi.org/10.1103/PhysRevLett.83.1566} {\bibfield  {journal} {\bibinfo
   {journal} {Phys. Rev. Lett.}\ }\textbf {\bibinfo {volume} {83}},\ \bibinfo
  {pages} {1566} (\bibinfo {year} {1999})}\BibitemShut {NoStop}%
\bibitem [{\citenamefont {Bedaque}\ \emph {et~al.}(2000)\citenamefont
  {Bedaque}, \citenamefont {Braaten},\ and\ \citenamefont
  {Hammer}}]{bedaque2000three}%
  \BibitemOpen
  \bibfield  {author} {\bibinfo {author} {\bibfnamefont {P.~F.}\ \bibnamefont
  {Bedaque}}, \bibinfo {author} {\bibfnamefont {E.}~\bibnamefont {Braaten}},\
  and\ \bibinfo {author} {\bibfnamefont {H.-W.}\ \bibnamefont {Hammer}},\
  }\bibfield  {title} {\bibinfo {title} {Three-body recombination in bose gases
  with large scattering length},\ }\href
  {https://doi.org/10.1103/PhysRevLett.85.908} {\bibfield  {journal} {\bibinfo
  {journal} {Phys. Rev. Lett.}\ }\textbf {\bibinfo {volume} {85}},\ \bibinfo
  {pages} {908} (\bibinfo {year} {2000})}\BibitemShut {NoStop}%
\bibitem [{\citenamefont {Braaten}\ and\ \citenamefont
  {Hammer}(2001)}]{braaten2001three}%
  \BibitemOpen
  \bibfield  {author} {\bibinfo {author} {\bibfnamefont {E.}~\bibnamefont
  {Braaten}}\ and\ \bibinfo {author} {\bibfnamefont {H.-W.}\ \bibnamefont
  {Hammer}},\ }\bibfield  {title} {\bibinfo {title} {Three-body recombination
  into deep bound states in a bose gas with large scattering length},\ }\href
  {https://doi.org/10.1103/PhysRevLett.87.160407} {\bibfield  {journal}
  {\bibinfo  {journal} {Phys. Rev. Lett.}\ }\textbf {\bibinfo {volume} {87}},\
  \bibinfo {pages} {160407} (\bibinfo {year} {2001})}\BibitemShut {NoStop}%
\bibitem [{\citenamefont {Hammer}\ \emph {et~al.}(2013)\citenamefont {Hammer},
  \citenamefont {Nogga},\ and\ \citenamefont
  {Schwenk}}]{HammerRevModPhys.85.197}%
  \BibitemOpen
  \bibfield  {author} {\bibinfo {author} {\bibfnamefont {H.-W.}\ \bibnamefont
  {Hammer}}, \bibinfo {author} {\bibfnamefont {A.}~\bibnamefont {Nogga}},\ and\
  \bibinfo {author} {\bibfnamefont {A.}~\bibnamefont {Schwenk}},\ }\bibfield
  {title} {\bibinfo {title} {Colloquium: Three-body forces: From cold atoms to
  nuclei},\ }\href {https://doi.org/10.1103/RevModPhys.85.197} {\bibfield
  {journal} {\bibinfo  {journal} {Rev. Mod. Phys.}\ }\textbf {\bibinfo {volume}
  {85}},\ \bibinfo {pages} {197} (\bibinfo {year} {2013})}\BibitemShut
  {NoStop}%
\bibitem [{\citenamefont {Zhu}\ and\ \citenamefont {Tan}(2017)}]{zhu2017three}%
  \BibitemOpen
  \bibfield  {author} {\bibinfo {author} {\bibfnamefont {S.}~\bibnamefont
  {Zhu}}\ and\ \bibinfo {author} {\bibfnamefont {S.}~\bibnamefont {Tan}},\
  }\bibfield  {title} {\bibinfo {title} {Three-body scattering hypervolumes of
  particles with short-range interactions},\ }\href@noop {} {\bibfield
  {journal} {\bibinfo  {journal} {arXiv preprint arXiv:1710.04147}\ } (\bibinfo
  {year} {2017})}\BibitemShut {NoStop}%
\bibitem [{\citenamefont {Braaten}\ and\ \citenamefont
  {Hammer}(2006)}]{braaten2006universality}%
  \BibitemOpen
  \bibfield  {author} {\bibinfo {author} {\bibfnamefont {E.}~\bibnamefont
  {Braaten}}\ and\ \bibinfo {author} {\bibfnamefont {H.-W.}\ \bibnamefont
  {Hammer}},\ }\bibfield  {title} {\bibinfo {title} {Universality in few-body
  systems with large scattering length},\ }\href
  {https://doi.org/https://doi.org/10.1016/j.physrep.2006.03.001} {\bibfield
  {journal} {\bibinfo  {journal} {Physics Reports}\ }\textbf {\bibinfo {volume}
  {428}},\ \bibinfo {pages} {259} (\bibinfo {year} {2006})}\BibitemShut
  {NoStop}%
\bibitem [{\citenamefont {Braaten}\ and\ \citenamefont
  {Nieto}(1999)}]{braaten1999quantum}%
  \BibitemOpen
  \bibfield  {author} {\bibinfo {author} {\bibfnamefont {E.}~\bibnamefont
  {Braaten}}\ and\ \bibinfo {author} {\bibfnamefont {A.}~\bibnamefont
  {Nieto}},\ }\bibfield  {title} {\bibinfo {title} {Quantum corrections to the
  energy density of a homogeneous bose gas},\ }\href
  {https://doi.org/10.1007/s100510050925} {\bibfield  {journal} {\bibinfo
  {journal} {The European Physical Journal B-Condensed Matter and Complex
  Systems}\ }\textbf {\bibinfo {volume} {11}},\ \bibinfo {pages} {143}
  (\bibinfo {year} {1999})}\BibitemShut {NoStop}%
\bibitem [{\citenamefont {Mestrom}\ \emph {et~al.}(2019)\citenamefont
  {Mestrom}, \citenamefont {Colussi}, \citenamefont {Secker},\ and\
  \citenamefont {Kokkelmans}}]{mestrom2019scattering}%
  \BibitemOpen
  \bibfield  {author} {\bibinfo {author} {\bibfnamefont {P.~M.~A.}\
  \bibnamefont {Mestrom}}, \bibinfo {author} {\bibfnamefont {V.~E.}\
  \bibnamefont {Colussi}}, \bibinfo {author} {\bibfnamefont {T.}~\bibnamefont
  {Secker}},\ and\ \bibinfo {author} {\bibfnamefont {S.~J. J. M.~F.}\
  \bibnamefont {Kokkelmans}},\ }\bibfield  {title} {\bibinfo {title}
  {Scattering hypervolume for ultracold bosons from weak to strong
  interactions},\ }\href {https://doi.org/10.1103/PhysRevA.100.050702}
  {\bibfield  {journal} {\bibinfo  {journal} {Phys. Rev. A}\ }\textbf {\bibinfo
  {volume} {100}},\ \bibinfo {pages} {050702} (\bibinfo {year}
  {2019})}\BibitemShut {NoStop}%
\bibitem [{\citenamefont {Mestrom}\ \emph {et~al.}(2020)\citenamefont
  {Mestrom}, \citenamefont {Colussi}, \citenamefont {Secker}, \citenamefont
  {Groeneveld},\ and\ \citenamefont {Kokkelmans}}]{mestrom2020van}%
  \BibitemOpen
  \bibfield  {author} {\bibinfo {author} {\bibfnamefont {P.~M.~A.}\
  \bibnamefont {Mestrom}}, \bibinfo {author} {\bibfnamefont {V.~E.}\
  \bibnamefont {Colussi}}, \bibinfo {author} {\bibfnamefont {T.}~\bibnamefont
  {Secker}}, \bibinfo {author} {\bibfnamefont {G.~P.}\ \bibnamefont
  {Groeneveld}},\ and\ \bibinfo {author} {\bibfnamefont {S.~J. J. M.~F.}\
  \bibnamefont {Kokkelmans}},\ }\bibfield  {title} {\bibinfo {title} {van der
  waals universality near a quantum tricritical point},\ }\href
  {https://doi.org/10.1103/PhysRevLett.124.143401} {\bibfield  {journal}
  {\bibinfo  {journal} {Phys. Rev. Lett.}\ }\textbf {\bibinfo {volume} {124}},\
  \bibinfo {pages} {143401} (\bibinfo {year} {2020})}\BibitemShut {NoStop}%
\bibitem [{\citenamefont {Wang}\ and\ \citenamefont
  {Tan}(2021)}]{wang2021threebody}%
  \BibitemOpen
  \bibfield  {author} {\bibinfo {author} {\bibfnamefont {Z.}~\bibnamefont
  {Wang}}\ and\ \bibinfo {author} {\bibfnamefont {S.}~\bibnamefont {Tan}},\
  }\bibfield  {title} {\bibinfo {title} {Three-body scattering hypervolume of
  particles with unequal masses},\ }\href
  {https://doi.org/10.1103/PhysRevA.103.063315} {\bibfield  {journal} {\bibinfo
   {journal} {Phys. Rev. A}\ }\textbf {\bibinfo {volume} {103}},\ \bibinfo
  {pages} {063315} (\bibinfo {year} {2021})}\BibitemShut {NoStop}%
\bibitem [{\citenamefont {Mestrom}\ \emph
  {et~al.}(2021{\natexlab{a}})\citenamefont {Mestrom}, \citenamefont {Colussi},
  \citenamefont {Secker}, \citenamefont {Li},\ and\ \citenamefont
  {Kokkelmans}}]{Mestrom2021pwave}%
  \BibitemOpen
  \bibfield  {author} {\bibinfo {author} {\bibfnamefont {P.~M.~A.}\
  \bibnamefont {Mestrom}}, \bibinfo {author} {\bibfnamefont {V.~E.}\
  \bibnamefont {Colussi}}, \bibinfo {author} {\bibfnamefont {T.}~\bibnamefont
  {Secker}}, \bibinfo {author} {\bibfnamefont {J.-L.}\ \bibnamefont {Li}},\
  and\ \bibinfo {author} {\bibfnamefont {S.~J. J. M.~F.}\ \bibnamefont
  {Kokkelmans}},\ }\bibfield  {title} {\bibinfo {title} {Three-body
  universality in ultracold $p$-wave resonant mixtures},\ }\href
  {https://doi.org/10.1103/PhysRevA.103.L051303} {\bibfield  {journal}
  {\bibinfo  {journal} {Phys. Rev. A}\ }\textbf {\bibinfo {volume} {103}},\
  \bibinfo {pages} {L051303} (\bibinfo {year}
  {2021}{\natexlab{a}})}\BibitemShut {NoStop}%
\bibitem [{\citenamefont {Mestrom}\ \emph
  {et~al.}(2021{\natexlab{b}})\citenamefont {Mestrom}, \citenamefont {Li},
  \citenamefont {Colussi}, \citenamefont {Secker},\ and\ \citenamefont
  {Kokkelmans}}]{mestrom2021threebody}%
  \BibitemOpen
  \bibfield  {author} {\bibinfo {author} {\bibfnamefont {P.~M.~A.}\
  \bibnamefont {Mestrom}}, \bibinfo {author} {\bibfnamefont {J.-L.}\
  \bibnamefont {Li}}, \bibinfo {author} {\bibfnamefont {V.~E.}\ \bibnamefont
  {Colussi}}, \bibinfo {author} {\bibfnamefont {T.}~\bibnamefont {Secker}},\
  and\ \bibinfo {author} {\bibfnamefont {S.~J. J. M.~F.}\ \bibnamefont
  {Kokkelmans}},\ }\bibfield  {title} {\bibinfo {title} {Three-body spin mixing
  in spin-1 bose-einstein condensates},\ }\href
  {https://doi.org/10.1103/PhysRevA.104.023321} {\bibfield  {journal} {\bibinfo
   {journal} {Phys. Rev. A}\ }\textbf {\bibinfo {volume} {104}},\ \bibinfo
  {pages} {023321} (\bibinfo {year} {2021}{\natexlab{b}})}\BibitemShut
  {NoStop}%
\bibitem [{\citenamefont {Esry}\ \emph {et~al.}(2001)\citenamefont {Esry},
  \citenamefont {Greene},\ and\ \citenamefont {Suno}}]{PhysRevA.65.010705}%
  \BibitemOpen
  \bibfield  {author} {\bibinfo {author} {\bibfnamefont {B.~D.}\ \bibnamefont
  {Esry}}, \bibinfo {author} {\bibfnamefont {C.~H.}\ \bibnamefont {Greene}},\
  and\ \bibinfo {author} {\bibfnamefont {H.}~\bibnamefont {Suno}},\ }\bibfield
  {title} {\bibinfo {title} {Threshold laws for three-body recombination},\
  }\href {https://doi.org/10.1103/PhysRevA.65.010705} {\bibfield  {journal}
  {\bibinfo  {journal} {Phys. Rev. A}\ }\textbf {\bibinfo {volume} {65}},\
  \bibinfo {pages} {010705} (\bibinfo {year} {2001})}\BibitemShut {NoStop}%
\bibitem [{\citenamefont {Yoshida}\ \emph {et~al.}(2018)\citenamefont
  {Yoshida}, \citenamefont {Saito}, \citenamefont {Waseem}, \citenamefont
  {Hattori},\ and\ \citenamefont {Mukaiyama}}]{PhysRevLett.120.133401}%
  \BibitemOpen
  \bibfield  {author} {\bibinfo {author} {\bibfnamefont {J.}~\bibnamefont
  {Yoshida}}, \bibinfo {author} {\bibfnamefont {T.}~\bibnamefont {Saito}},
  \bibinfo {author} {\bibfnamefont {M.}~\bibnamefont {Waseem}}, \bibinfo
  {author} {\bibfnamefont {K.}~\bibnamefont {Hattori}},\ and\ \bibinfo {author}
  {\bibfnamefont {T.}~\bibnamefont {Mukaiyama}},\ }\bibfield  {title} {\bibinfo
  {title} {Scaling law for three-body collisions of identical fermions with
  $p$-wave interactions},\ }\href
  {https://doi.org/10.1103/PhysRevLett.120.133401} {\bibfield  {journal}
  {\bibinfo  {journal} {Phys. Rev. Lett.}\ }\textbf {\bibinfo {volume} {120}},\
  \bibinfo {pages} {133401} (\bibinfo {year} {2018})}\BibitemShut {NoStop}%
\bibitem [{\citenamefont {Çağrı Top}\ \emph {et~al.}(2020)\citenamefont
  {Çağrı Top}, \citenamefont {Margalit},\ and\ \citenamefont
  {Ketterle}}]{top2020spinpolarized}%
  \BibitemOpen
  \bibfield  {author} {\bibinfo {author} {\bibfnamefont {F.}~\bibnamefont
  {Çağrı Top}}, \bibinfo {author} {\bibfnamefont {Y.}~\bibnamefont
  {Margalit}},\ and\ \bibinfo {author} {\bibfnamefont {W.}~\bibnamefont
  {Ketterle}},\ }\href@noop {} {\bibinfo {title} {Spin-polarized fermions with
  $p$-wave interactions}} (\bibinfo {year} {2020}),\ \Eprint
  {https://arxiv.org/abs/2009.05913} {arXiv:2009.05913 [cond-mat.quant-gas]}
  \BibitemShut {NoStop}%
\bibitem [{\citenamefont {Nielsen}\ \emph {et~al.}(2001)\citenamefont
  {Nielsen}, \citenamefont {Fedorov}, \citenamefont {Jensen},\ and\
  \citenamefont {Garrido}}]{nielsen2001three}%
  \BibitemOpen
  \bibfield  {author} {\bibinfo {author} {\bibfnamefont {E.}~\bibnamefont
  {Nielsen}}, \bibinfo {author} {\bibfnamefont {D.}~\bibnamefont {Fedorov}},
  \bibinfo {author} {\bibfnamefont {A.}~\bibnamefont {Jensen}},\ and\ \bibinfo
  {author} {\bibfnamefont {E.}~\bibnamefont {Garrido}},\ }\bibfield  {title}
  {\bibinfo {title} {The three-body problem with short-range interactions},\
  }\href {https://doi.org/https://doi.org/10.1016/S0370-1573(00)00107-1}
  {\bibfield  {journal} {\bibinfo  {journal} {Physics Reports}\ }\textbf
  {\bibinfo {volume} {347}},\ \bibinfo {pages} {373} (\bibinfo {year}
  {2001})}\BibitemShut {NoStop}%
\bibitem [{\citenamefont {Hammer}\ and\ \citenamefont
  {Lee}(2010)}]{hammer2010causality}%
  \BibitemOpen
  \bibfield  {author} {\bibinfo {author} {\bibfnamefont {H.-W.}\ \bibnamefont
  {Hammer}}\ and\ \bibinfo {author} {\bibfnamefont {D.}~\bibnamefont {Lee}},\
  }\bibfield  {title} {\bibinfo {title} {Causality and the effective range
  expansion},\ }\href
  {https://doi.org/https://doi.org/10.1016/j.aop.2010.06.006} {\bibfield
  {journal} {\bibinfo  {journal} {Annals of Physics}\ }\textbf {\bibinfo
  {volume} {325}},\ \bibinfo {pages} {2212} (\bibinfo {year}
  {2010})}\BibitemShut {NoStop}%
\bibitem [{\citenamefont {Hammer}\ and\ \citenamefont
  {Lee}(2009)}]{hammer2009}%
  \BibitemOpen
  \bibfield  {author} {\bibinfo {author} {\bibfnamefont {H.-W.}\ \bibnamefont
  {Hammer}}\ and\ \bibinfo {author} {\bibfnamefont {D.}~\bibnamefont {Lee}},\
  }\bibfield  {title} {\bibinfo {title} {Causality and universality in
  low-energy quantum scattering},\ }\href
  {https://doi.org/https://doi.org/10.1016/j.physletb.2009.10.033} {\bibfield
  {journal} {\bibinfo  {journal} {Physics Letters B}\ }\textbf {\bibinfo
  {volume} {681}},\ \bibinfo {pages} {500} (\bibinfo {year}
  {2009})}\BibitemShut {NoStop}%
\bibitem [{\citenamefont {Huang}\ and\ \citenamefont
  {Yang}(1957)}]{huang1957quantum}%
  \BibitemOpen
  \bibfield  {author} {\bibinfo {author} {\bibfnamefont {K.}~\bibnamefont
  {Huang}}\ and\ \bibinfo {author} {\bibfnamefont {C.~N.}\ \bibnamefont
  {Yang}},\ }\bibfield  {title} {\bibinfo {title} {Quantum-mechanical many-body
  problem with hard-sphere interaction},\ }\href
  {https://doi.org/10.1103/PhysRev.105.767} {\bibfield  {journal} {\bibinfo
  {journal} {Phys. Rev.}\ }\textbf {\bibinfo {volume} {105}},\ \bibinfo {pages}
  {767} (\bibinfo {year} {1957})}\BibitemShut {NoStop}%
\bibitem [{\citenamefont {Lee}\ \emph {et~al.}(1957)\citenamefont {Lee},
  \citenamefont {Huang},\ and\ \citenamefont {Yang}}]{lee1957eigenvalues}%
  \BibitemOpen
  \bibfield  {author} {\bibinfo {author} {\bibfnamefont {T.~D.}\ \bibnamefont
  {Lee}}, \bibinfo {author} {\bibfnamefont {K.}~\bibnamefont {Huang}},\ and\
  \bibinfo {author} {\bibfnamefont {C.~N.}\ \bibnamefont {Yang}},\ }\bibfield
  {title} {\bibinfo {title} {Eigenvalues and eigenfunctions of a bose system of
  hard spheres and its low-temperature properties},\ }\href
  {https://doi.org/10.1103/PhysRev.106.1135} {\bibfield  {journal} {\bibinfo
  {journal} {Phys. Rev.}\ }\textbf {\bibinfo {volume} {106}},\ \bibinfo {pages}
  {1135} (\bibinfo {year} {1957})}\BibitemShut {NoStop}%
\bibitem [{\citenamefont {Roth}\ and\ \citenamefont
  {Feldmeier}(2001)}]{roth2001effective}%
  \BibitemOpen
  \bibfield  {author} {\bibinfo {author} {\bibfnamefont {R.}~\bibnamefont
  {Roth}}\ and\ \bibinfo {author} {\bibfnamefont {H.}~\bibnamefont
  {Feldmeier}},\ }\bibfield  {title} {\bibinfo {title} {Effective s- and p-wave
  contact interactions in trapped degenerate fermi gases},\ }\href
  {https://doi.org/10.1103/PhysRevA.64.043603} {\bibfield  {journal} {\bibinfo
  {journal} {Phys. Rev. A}\ }\textbf {\bibinfo {volume} {64}},\ \bibinfo
  {pages} {043603} (\bibinfo {year} {2001})}\BibitemShut {NoStop}%
\bibitem [{\citenamefont {Kanjilal}\ and\ \citenamefont
  {Blume}(2004)}]{kanjilal2004nondivergent}%
  \BibitemOpen
  \bibfield  {author} {\bibinfo {author} {\bibfnamefont {K.}~\bibnamefont
  {Kanjilal}}\ and\ \bibinfo {author} {\bibfnamefont {D.}~\bibnamefont
  {Blume}},\ }\bibfield  {title} {\bibinfo {title} {Nondivergent
  pseudopotential treatment of spin-polarized fermions under one- and
  three-dimensional harmonic confinement},\ }\href
  {https://doi.org/10.1103/PhysRevA.70.042709} {\bibfield  {journal} {\bibinfo
  {journal} {Phys. Rev. A}\ }\textbf {\bibinfo {volume} {70}},\ \bibinfo
  {pages} {042709} (\bibinfo {year} {2004})}\BibitemShut {NoStop}%
\bibitem [{\citenamefont {Derevianko}(2005)}]{derevianko2005revised}%
  \BibitemOpen
  \bibfield  {author} {\bibinfo {author} {\bibfnamefont {A.}~\bibnamefont
  {Derevianko}},\ }\bibfield  {title} {\bibinfo {title} {Revised huang-yang
  multipolar pseudopotential},\ }\href
  {https://doi.org/10.1103/PhysRevA.72.044701} {\bibfield  {journal} {\bibinfo
  {journal} {Phys. Rev. A}\ }\textbf {\bibinfo {volume} {72}},\ \bibinfo
  {pages} {044701} (\bibinfo {year} {2005})}\BibitemShut {NoStop}%
\bibitem [{\citenamefont {Stock}\ \emph {et~al.}(2005)\citenamefont {Stock},
  \citenamefont {Silberfarb}, \citenamefont {Bolda},\ and\ \citenamefont
  {Deutsch}}]{stock2005generalized}%
  \BibitemOpen
  \bibfield  {author} {\bibinfo {author} {\bibfnamefont {R.}~\bibnamefont
  {Stock}}, \bibinfo {author} {\bibfnamefont {A.}~\bibnamefont {Silberfarb}},
  \bibinfo {author} {\bibfnamefont {E.~L.}\ \bibnamefont {Bolda}},\ and\
  \bibinfo {author} {\bibfnamefont {I.~H.}\ \bibnamefont {Deutsch}},\
  }\bibfield  {title} {\bibinfo {title} {Generalized pseudopotentials for
  higher partial wave scattering},\ }\href
  {https://doi.org/10.1103/PhysRevLett.94.023202} {\bibfield  {journal}
  {\bibinfo  {journal} {Phys. Rev. Lett.}\ }\textbf {\bibinfo {volume} {94}},\
  \bibinfo {pages} {023202} (\bibinfo {year} {2005})}\BibitemShut {NoStop}%
\bibitem [{\citenamefont {Pricoupenko}(2006)}]{pricoupenko2006modeling}%
  \BibitemOpen
  \bibfield  {author} {\bibinfo {author} {\bibfnamefont {L.}~\bibnamefont
  {Pricoupenko}},\ }\bibfield  {title} {\bibinfo {title} {Modeling interactions
  for resonant $p$-wave scattering},\ }\href
  {https://doi.org/10.1103/PhysRevLett.96.050401} {\bibfield  {journal}
  {\bibinfo  {journal} {Phys. Rev. Lett.}\ }\textbf {\bibinfo {volume} {96}},\
  \bibinfo {pages} {050401} (\bibinfo {year} {2006})}\BibitemShut {NoStop}%
\bibitem [{\citenamefont {Idziaszek}\ and\ \citenamefont
  {Calarco}(2006)}]{idziaszek2006pseudopotential}%
  \BibitemOpen
  \bibfield  {author} {\bibinfo {author} {\bibfnamefont {Z.}~\bibnamefont
  {Idziaszek}}\ and\ \bibinfo {author} {\bibfnamefont {T.}~\bibnamefont
  {Calarco}},\ }\bibfield  {title} {\bibinfo {title} {Pseudopotential method
  for higher partial wave scattering},\ }\href
  {https://doi.org/10.1103/PhysRevLett.96.013201} {\bibfield  {journal}
  {\bibinfo  {journal} {Phys. Rev. Lett.}\ }\textbf {\bibinfo {volume} {96}},\
  \bibinfo {pages} {013201} (\bibinfo {year} {2006})}\BibitemShut {NoStop}%
\bibitem [{\citenamefont {Reichenbach}\ \emph {et~al.}(2006)\citenamefont
  {Reichenbach}, \citenamefont {Silberfarb}, \citenamefont {Stock},\ and\
  \citenamefont {Deutsch}}]{reichenbach2006quasi}%
  \BibitemOpen
  \bibfield  {author} {\bibinfo {author} {\bibfnamefont {I.}~\bibnamefont
  {Reichenbach}}, \bibinfo {author} {\bibfnamefont {A.}~\bibnamefont
  {Silberfarb}}, \bibinfo {author} {\bibfnamefont {R.}~\bibnamefont {Stock}},\
  and\ \bibinfo {author} {\bibfnamefont {I.~H.}\ \bibnamefont {Deutsch}},\
  }\bibfield  {title} {\bibinfo {title} {Quasi-hermitian pseudopotential for
  higher partial wave scattering},\ }\href
  {https://doi.org/10.1103/PhysRevA.74.042724} {\bibfield  {journal} {\bibinfo
  {journal} {Phys. Rev. A}\ }\textbf {\bibinfo {volume} {74}},\ \bibinfo
  {pages} {042724} (\bibinfo {year} {2006})}\BibitemShut {NoStop}%
\bibitem [{\citenamefont {Idziaszek}(2009)}]{idziaszek2009analytical}%
  \BibitemOpen
  \bibfield  {author} {\bibinfo {author} {\bibfnamefont {Z.}~\bibnamefont
  {Idziaszek}},\ }\bibfield  {title} {\bibinfo {title} {Analytical solutions
  for two atoms in a harmonic trap: $p$-wave interactions},\ }\href
  {https://doi.org/10.1103/PhysRevA.79.062701} {\bibfield  {journal} {\bibinfo
  {journal} {Phys. Rev. A}\ }\textbf {\bibinfo {volume} {79}},\ \bibinfo
  {pages} {062701} (\bibinfo {year} {2009})}\BibitemShut {NoStop}%
\bibitem [{\citenamefont {Jona-Lasinio}\ \emph {et~al.}(2008)\citenamefont
  {Jona-Lasinio}, \citenamefont {Pricoupenko},\ and\ \citenamefont
  {Castin}}]{castin2008threepolarized}%
  \BibitemOpen
  \bibfield  {author} {\bibinfo {author} {\bibfnamefont {M.}~\bibnamefont
  {Jona-Lasinio}}, \bibinfo {author} {\bibfnamefont {L.}~\bibnamefont
  {Pricoupenko}},\ and\ \bibinfo {author} {\bibfnamefont {Y.}~\bibnamefont
  {Castin}},\ }\bibfield  {title} {\bibinfo {title} {Three fully polarized
  fermions close to a $\mathit{p}$-wave feshbach resonance},\ }\href
  {https://doi.org/10.1103/PhysRevA.77.043611} {\bibfield  {journal} {\bibinfo
  {journal} {Phys. Rev. A}\ }\textbf {\bibinfo {volume} {77}},\ \bibinfo
  {pages} {043611} (\bibinfo {year} {2008})}\BibitemShut {NoStop}%
\end{thebibliography}%

\end{document}